\documentclass[12pt,preprint]{aastex}

\shorttitle{Warm Dust Emission in z$\sim$6 Quasars}
\shortauthors{Wang et al.}

\begin{document}

\title{Thermal Emission from Warm Dust in the Most Distant Quasars}

\author{Ran Wang\altaffilmark{1,2},
Chris L. Carilli\altaffilmark{2},
Jeff Wagg\altaffilmark{2}
Frank Bertoldi\altaffilmark{3},
Fabian Walter\altaffilmark{4},
Karl M. Menten\altaffilmark{5},
Alain Omont\altaffilmark{6},
Pierre Cox\altaffilmark{7},
Michael A. Strauss\altaffilmark{8},
Xiaohui Fan\altaffilmark{9},
Linhua Jiang\altaffilmark{9}
Donald P. Schneider\altaffilmark{10}}
\altaffiltext{1}{Department of Astronomy, Peking University, Beijing 100871, China}
\altaffiltext{2}{National Radio Astronomy Observatory, PO Box 0, Socorro, NM, USA 87801}
\altaffiltext{3}{Argelander-Institut f$\rm \ddot u$r Astronomie, University of Bonn, Auf dem H$\rm \ddot u$gel 71, 53121 Bonn, Germany}
\altaffiltext{4}{Max-Planck-Institute for Astronomy, K$\rm \ddot o$nigsstuhl 17, 69117 Heidelberg, Germany}
\altaffiltext{5}{Max-Planck-Institut f$\rm \ddot u$r Radioastronomie, Auf dem H$\rm \ddot u$gel 71, 53121 Bonn, Germany}
\altaffiltext{6}{Institut d'Astrophysique de Paris, CNRS and Universite Pierre et Marie Curie, Paris, France}
\altaffiltext{7}{Institute de Radioastronomie Millimetrique, St. Martin d'Heres, F-38406, France}
\altaffiltext{8}{Department of Astrophysical Sciences, Princeton University, Princeton, NJ, USA, 08544}
\altaffiltext{9}{Steward Observatory, The University of Arizona, Tucson, AZ 85721}
\altaffiltext{10}{Department of Astronomy and Astrophysics, The Pennsylvania State University, University Park, PA 16802}
\begin{abstract}

We report new continuum observations of fourteen z$\sim$6 
quasars at 250 GHz and fourteen quasars at 1.4 GHz. We  
summarize all recent millimeter and radio observations of 
the sample of the thirty-three quasars known with $\rm 5.71\le z\le 6.43$, 
and present a study of the rest frame far-infrared (FIR) 
properties of this sample. These quasars were observed with 
the Max Plank Millimeter Bolometer Array (MAMBO) at 250 GHz 
with mJy sensitivity, and $\rm 30\%$ of them were detected. 
We also recover the average 250 GHz flux density of the MAMBO 
undetected sources at $\rm 4\sigma$, by stacking the on-source 
measurements. The derived mean radio-to-UV spectral energy distributions (SEDs) 
of the full sample and the 250 GHz non-detections show no significant 
difference from that of lower-redshift 
optical quasars. Obvious FIR excesses are seen in 
the individual SEDs of the strong 250 GHz detections, with FIR-to-radio 
emission ratios consistent with that of typical star forming 
galaxies. Most 250 GHz-detected sources follow the 
$\rm L_{FIR}$--$\rm L_{bol}$ relationship derived from a sample 
of local IR luminous quasars ($\rm L_{IR}>10^{12}L_{\odot}$), 
while the average $\rm L_{FIR}/L_{bol}$ ratio of the non-detections 
is consistent with that of the optically-selected PG quasars. 
The MAMBO detections also tend to have weaker $\rm Ly\alpha$ 
emission than the non-detected sources. We discuss possible 
FIR dust heating sources, and critically assess the possibility 
of active star formation in the host galaxies of the z$\sim$6 
quasars. The average star formation rate of the MAMBO non-detections is 
likely to be less than a few hundred $\rm M_{\odot}\,yr^{-1}$, 
but in the strong detections, the host galaxy star formation is probably at a rate of 
$\rm \gtrsim10^{3}\,M_{\odot}\,yr^{-1}$, which dominates the FIR 
dust heating.

\end{abstract}


\section{Introduction}

More than thirty quasars at z$\sim$6 have been discovered 
(e.g., Fan et al. 2000, 2001a, 2003, 2004, 2006a; Jiang
et al. 2007a; Willott et al. 2007). These objects are
the earliest massive black hole accretion
systems known (Jiang et al. 2006; 2007b), seen at an epoch
close to the end of cosmic reionization (Fan et al. 2006b).
They provide us with unique information on both the growth
of supermassive black holes (SMBHs) and the formation
of massive galaxies when the age of the universe was 
$\rm \lesssim1\,Gyr$.

There are fundamental relationships between SMBH mass and bulge stellar 
mass/velocity dispersion in the local universe (e.g., Tremaine et
al. 2002; Marconi \& Hunt 2003; Hopkins et al. 2007), indicating that
the formation of SMBHs and their spheroidal hosts are
coupled. 
Active galactic nuclei (AGNs) buried in dusty starburst environments have been discovered in  
samples of low redshift ultraluminous infrared galaxies (ULIRGs), 
which are believed to be a transition phase between starburst 
mergers and typical optically bright AGNs (Sanders et al. 1988; 
Wu et al. 1998; Zheng et al. 2002). 
Studies of large samples of galaxies and AGNs 
also provide clues about the 'downsizing' process in galaxy and SMBH formation, 
whereby the formation of the most massive systems occur at early epochs
(Cowie et al. 1996; Heckman et al. 2004; Kauffmann \& Heckman 2005).

Examples of massive quasars hosted by interacting systems or 
bright submillimeter galaxies have been found at redshifts 
greater than 4, such as BR 1202-0725 at z=4.7 and BRI1335-0417
at z=4.4 (Omont et al. 1996a, 1996b; Benford et al.
1999; Beelen et al. 2006). These objects are all characterized by
strong molecular CO line emission (Guilloteau et al. 1997, 1999; Carilli
et al. 2002; Solomon \& Vanden Bout 2005; Riechers et al. 2006) and
FIR (Guilloteau et al. 1999) and radio (Momjian et al.
2005, 2007) emission originating from a starburst, with implied star 
formation rates of a few thousand $\rm M_{\odot}\,yr^{-1}$. 
The results suggest that the galaxies are at an earlier evolution 
stage than are typical optically bright quasars in which the central 
AGN dominates the emission from X-ray to radio.

These studies have been extended to z$\sim$6 quasar sample with
sensitive submillimeter, millimeter, and radio telescopes (Priddey
et al. 2003b; Bertoldi et al. 2003; Petric et al. 2003; Carilli et al. 2004; Wang et al.
2007). 
About 1/3 of these z$\sim$6 quasars
were detected at millimeter wavelengths, 
at mJy sensitivity (Priddey et al. 2003b; Wang et al. 2007). The 
millimeter detections imply FIR luminosities of $\rm 10^{12}\sim10^{13}\,L_{\odot}$
and dust masses of $\rm \gtrsim10^{8}\,M_{\odot}$ in the quasar host galaxies
(Bertoldi et al. 2003a, Beelen et al. 2006). Such huge dust masses
require rapid metal and dust enrichment via a starburst in the 
early evolution of these galaxies (Bertoldi et al. 2003a; 
Walter et al. 2003; Venkatesan et al. 2006; Li et al. 2007b; Dwek et al. 2007).

The heating sources of the FIR-emitting dust in the quasar
systems at z$\sim$6 have become a key question. The reprocessed emission from 
star formation-heated dust can provide a direct estimate   
of the star formation rate, thus constraining the bulge building 
stage in these quasar hosts. 
Observations at submillimeter wavelengths imply that 
the FIR emission in the two sources with the strongest millimeter detections, 
SDSS J114816.64+525150.3 at z=6.42 (hereafter J1148+5251) 
and SDSS J092721.82+200123.7 at z=5.77 (hereafter J0927+2001), 
is from dust components with 
temperatures of 50 to 60 K (Beelen et al. 2006; Wang et al. 2008a).
Large amounts of highly excited molecular CO were also detected 
in the host galaxies of these two sources (Bertoldi et al. 2003b; 
Walter et al. 2003; Carilli et al. 2007), as well as strong [C {\small II}] 
158$\,\mu$m ISM gas cooling line emission in J1148+5251 (Maiolino et al. 2005).
These results suggest that star formation plays an important 
role in the heating of FIR-emitting warm dust. The implied star 
formation rate is $\rm \gtrsim10^{3}\,M_{\odot}\,yr^{-1}$, which argues 
for active bulge building in these two z$\sim$6 quasars. 

We have a long standing program to study the dust and gas emission
from the z$\sim$6 quasars (e.g., Wang et al. 2007, 2008a). In this paper, we 
report new observations and present an FIR emission study 
of the full sample of thirty-three z$\sim$6 quasars discovered to date, 
aiming (i) to give a general view of the FIR emission properties of the sample, and
(ii) to further constrain the dust heating and star
forming activity in the quasar hosts. We describe the full 
quasar sample, the new observations, and summarize the current millimeter and
radio results in section 2. The analysis of the full sample is given in Sections 3, 4, and 5.
We present a discussion of star formation in Section 6,
and give a brief summary in Section 7. We adopt a $\rm \Lambda$-model
cosmology with $\rm H_{0}=71km\ s^{-1}\ Mpc^{-1}$, $\rm
\Omega_{M}=0.27$ and $\rm \Omega_{\Lambda}=0.73$ throughout this
paper (Spergel et al. 2007).

\section{Sample and observations}

\subsection{The sample of z$\sim$6 quasars}

Thirty-three quasars have been discovered at z$\sim$6 
($\rm 5.71 \le z \le 6.43$, see Table 1 and 2). Twenty-two 
of these objects were selected from $\sim$8000 deg$^{2}$ 
of the Sloan Digital Sky Survey (SDSS, York et al. 2000) imaging data 
(Fan et al. 2000, 2001a, 2003, 2004, 2006a, 2008, in prep.; 
Goto 2006). These objects typically have 
rest frame $\rm 1450\AA$ AB magnitudes $\rm m_{1450}<20$.
The SDSS imaging survey is using 
a dedicated 2.5m telescope and a large format CCD camera 
(Gunn et al. 2006) 
at the Apache Point Observatory in New Mexico
to obtain images in five broad bands ($\rm u$, $\rm g$, $\rm r$, $\rm i$ and $\rm z$,
centered at 3551, 4686, 6166, 7480 and 8932 $\rm \AA$, respectively; 
Fukugita et al. 1996) 
of high Galactic latitude sky in the Northern Galactic Cap.
The imaging data are processed by the astrometric 
pipeline (Pier et al. 2003) and photometric pipeline (Lupton et al. 2001), 
resulting in 
astrometric calibration errors of $\rm < 0.1''$ rms per coordinate, and
photometric calibration to better than 0.03 mag
(Hogg et al. 2001, Smith et al. 2002, Ivezi{\' c} et al. 2004, Tucker 
et al. 2006).
These data have been made available to the public 
in a series of data releases (EDR: Stoughton et al. 2002; DR1 
-- DR6:
Abazajian et al. 2003, 2004, 2005, Adelman-McCarthy et al. 2006, 
2007, 2008).

Five additional z$\sim$6 quasars were discovered in the SDSS
Deep Equatorial Stripe ('Stripe 82') by Jiang et al. (2007a). These five
sources were selected from deep optical imaging of 
260 deg$^2$. These objects are about
one magnitude fainter in the optical (i.e., $\rm 20<m_{1450}<21.5$) than the
primary SDSS sample.
One source in this sample, J020332.39+001229.3 (hereafter J0203+0012), was also
independently discovered in the UKIRT Infrared
Deep Sky Survey (Venemans et al. 2007).

Four new quasars at
$\rm z>6$ have been published (Willott et al. 2007)
from the Canada-France High-z Quasar
Survey (CFHQS). These quasars were selected from
imaging of 400 deg$^2$; three of them have
$\rm m_{1450}>21$. The fourth source, CFHQS J1509-1749,
is comparable in brightness to the faintest source in
the primary SDSS sample, but the optical spectrum
is much steeper than that of the typical quasars,
indicating significant dust reddening.

Another two z$\sim$6 quasars were discovered in the NOAO Deep Wide-Field
Survey (Jannuzi et al. 2004) using optical, infrared, and radio data 
(McGreer et al. 2006; Cool et al. 2006). One of them, J1427+3312 at z=6.12,
was selected by McGreer et al. (2006) from 4 deg$^{2}$ 
by matching infrared and radio sources to optical counterparts.
It is the brightest radio source in the z$\sim$6 quasar sample. The
UV to optical spectrum of this object is likely to be reddened by
dust, and the rest frame $\rm 1450\AA$ magnitude derived from near-infrared
data is $\rm m_{1450}=20.33$\footnote{derived from the absolute 
magnitude of $\rm M_{1450}=-26.4$ presented in McGreen et al. (2006)}. 
The other quasar was infrared selected from Spitzer mid-infrared 
data, and identified in the AGN and Galaxy Evolution Survey 
(Cool et al. 2006) with $\rm m_{1450}=20.62$.  

\subsection{New observations}

Twenty-two sources in the z$\sim$6 quasar sample have published 250 GHz 
continuum observations with the Max-Planck 
Millimeter Bolometer Array (MAMBO, Kreysa et al. 1998) on the IRAM 30-m
telescope (Bertoldi et al. 2003a; Petric et al. 2003; Wang et al. 2007; 
Willott et al. 2007; Priddey et al. 2008), and nineteen sources in the 
sample have published 1.4 GHz continuum observations with the VLA (Petric et al. 2003; 
Carilli et al. 2004; Wang et al. 2007; McGreer et al. 2006). 
We report our new MAMBO and VLA observations of totally nineteen z$\sim$6 quasars in this paper. 
These sources are unobserved or marginally detected in the previous 
observations, and we present new data for them at either 250 GHz or 1.4 GHz.
We describe the new observations below and list the measurements for the 
nineteen quasars in Table 1. We also summarise 
the previous observations for other z$\sim$6 quasars in Table 2.

We present new 250 GHz observations for fourteen z$\sim$6 quasars, including 
eleven sources that have never been previously observed. 
The new observations have been obtained with the MAMBO-II 117-element 
array in Winter 2006-2007, we adopted the same observing
mode as was used in the previous observations (Wang et al. 2007),
i.e., doing chopping photometry at 2 Hz with $\rm 32''$
throw in azimuth. We typically spent 1-3 hours on each source to
achieve the sub-mJy rms noise level reached in previous observations.
The data were reduced with the standard MOPSIC
pipeline (Zylka 1998). Three sources, J1044-0125,
J1425+3254, and J2054-0005, were detected from
the new data at $\rm >4\sigma$ level. The MAMBO  
detection of J1044-0125 was recently reported in 
Priddey et al. (2008) from an independent observation 
with a 250 GHz flux density of $\rm 2.5\pm0.6$. 
We observed this source down to a lower rms noise level, 
and our result, $\rm 1.8\pm0.4\,mJy$, is a bit lower compared to the value in 
Priddey et al. (2008) but consistent within the error limits (see Table 1).  
By including the data from the new observations, we also 
confirmed the previous $\rm 3\sigma$ detection in 
the source J0818+1722 (Wang et al. 2007) with a 
lower rms of 0.4 mJy.

New 1.4 GHz data were obtained with the VLA for fourteen 
quasars, including thirteen previously unobserved 
sources and a published marginally detected source, 
J0927+2001 ($\rm 3.2\sigma$, Wang et al. 2007).
The new observations were carried out in 2006 and 2007 with the array in the A 
or B configurations, down to an rms noise level of $\rm \sim20\,\mu Jy$
for most of the sources. There is a $\sim$1 Jy source in the
field of the quasar J0303-0019, leading to a high rms
of $\rm 66\,\mu Jy$ in the image. All the data
were reduced using AIPS.
One source, J0203+0012, was detected by our new observation with a 1.4 GHz flux density 
of $\rm S_{\nu}=195\pm22\,\mu Jy$, and the previous detection of J0927+2001 
was confirmed at $\rm \gtrsim4\sigma$ level. 

There are fairly bright and extended radio sources in the 
1.4 GHz field images of the CFHQS quasars J0033-0125 and J1641+3755 (Figure 1), though these two quasars are not detected. 
One double-lobe radio source is found about $\rm9'$ away from the optical position 
of the  quasar J0033-0125, with integrated 1.4 GHz flux density of $\rm \sim23\,mJy$. This radio source was also 
detected in the NVSS survey (Condon et al. 1998) and FIRST survey (White et al. 1997), and 
there is no published deep optical data for this area yet. 
Five radio sources with integrated flux densities $\rm >10\,mJy$ are seen 
in the field of the source J1641+3755, including extended 
double radio sources, within 10$\rm '$ from the optical quasar position. 
According to the FIRST 1.4 GHz source counts (White et al. 1997),
about 1.6 source with $\rm f_{1.4GHz}>10\,mJy$ is expected within this amount of sky area.
The brightest radio source in the field of J1641+3755 was 
identified as a radio-loud AGN with a spectroscopic redshift of z=0.162 (Best et al. 2005).
Further optical imaging and spectrography will address the question of whether the
bright radio sources in the fields of the two CFHQS quasars is due to foreground  
galaxy clusters. We notice that bright extended radio sources and overdensities were previously reported in the 1.4 GHz 
observations of the two SDSS z$\sim$6 quasars J1030+0524 
and J1148+5251 (Petric et al. 2003; Carilli et al. 2004). Carilli et al. (2004) 
investigated the SDSS data of the J1148+5251 field, and however, there is no evidence for 
dense foreground clusters or gravitational lensing. 

\subsection{Summary of the millimeter and radio results}

In summary, all the thirty-three source in the z$\sim$6 quasar 
sample have been observed by MAMBO at 250 GHz. The observations 
have rms values in the range of $\rm 0.4-1.1\,mJy$ with a median value of about $\rm
0.6\,mJy$, and ten sources are detected (see Table 1 and 2).
This yields a detection rate of $\rm 30\pm10\%$ which is consistent 
with the submillimeter and millimeter detection rates of optically bright quasars at redshifts 2 and 4 (Priddey et al. 2003a; Omont
et al. 2001; 2003; Carilli et al. 2001).

Thirty-two z$\sim$6 quasars 
have 1.4 GHz radio continuum observations by the VLA, with a median rms level of $\rm 20\,\mu Jy$, 
Ten out of the thirty-two sources have been detected, including four 250 GHz detections (see Table 1 and 2). 
Two of the radio detections, J0836+0054 and J1427+3312, have 
flux densities $\rm >1mJy$. These two sources, together with the new radio detection J0203+0012,  
have radio loudness\footnote{The $\rm 4400\AA$ and 5 GHz flux densities are derived
from the 1450$\rm \AA$ magnitude and the observed 1.4 GHz flux density, 
assuming power law spectra indices of -0.5 in the optical 
and -0.75 in the radio, respectively.} 
$\rm R\equiv f_{\nu,5GHz}/f_{\nu,4400}\ge10$,
and hence are radio loud according to the
definition by Kellermann et al. (1989).

Data at other wavelengths for the analysis in this
paper were collected from the literature, including optical data
from the original papers (Fan et al. 2000; 2001a; 2003; 2004; 2006a;
2008, in prep; Goto 2006; McGreer et al. 2006; Willott et al. 2007;
Jiang et al. 2008), near-infrared data from Jiang et al. (2006),
submillimeter data from Priddey et al. (2003b), Robson et al. (2004) and Beleen et al. (2006),
and the millimeter continuum data from the PdBI CO observations 
(Carilli et al. 2007; Wang et al. 2008b, in prep.).

\section{The UV-to-radio SEDs of the z$\sim$6 quasars}

Using the 250 GHz and 1.4 GHz continuum measurements, 
we derive the average FIR and radio emission of these 
z$\sim$6 quasars and compare it to the typical 
quasars at low redshift. The calculations
are performed for three groups: (I) the whole
sample of thirty-three sources, (II) the subsample of
ten sources detected with MAMBO at $\rm \ge3\sigma$, and 
(III) the subsample of twenty-three sources not detected with MAMBO.
Using the MAMBO 250GHz data, we calculate the weighted
average 250GHz flux densities for each group, namely:  \\ 
\begin{equation}
\rm \left\langle f_{250GHz}\right\rangle =\frac{\sum w_{i}f_{250GHz,i}}{\sum w_{i}}
\end{equation}
\begin{equation}
\rm w_{i}=\frac{1}{\sigma^{2}_{250GHz,i}}
\end{equation}
\begin{equation}
\rm \left\langle \sigma_{250GHz}\right\rangle=\left(\frac{\sum \left(w_{i}\sigma_{250GHz,i}\right)^2}{(\sum w_{i})^2}\right)^{0.5}
\end{equation} 
where $\rm f_{250GHz,i}$ is the MAMBO observed flux density for each source, 
and $\sigma_{250GHz,i}$ is the $\rm 1\sigma$ rms. 
For the calculation of $\rm \langle f_{250GHz}\rangle$ of the whole sample and the  
subsample of 250 GHz non-detections, we weighed the data by $\rm 1/\sigma^{2}_{250GHz,i}$ (equation (2)).
However, equal weights were adopted (i.e. $\rm w_{i}=1$) when calculate $\rm \langle f_{250GHz}\rangle$ 
for the 250 GHz detections, as the observed rms values spread in a smaller range (0.36$\sim$0.75 mJy) 
and we do not want to bias the average toward the weakest sources.
The average flux density ratios between 250GHz and 
the rest frame 1450$\rm \AA$ are then calculated 
as $\rm \langle f_{250GHz}\rangle/\langle f_{1450}\rangle$, where the average 1450$\rm \AA$ 
flux density $\rm \langle f_{1450}\rangle$ is derived from the AB magnitudes presented 
in the discovering papers (see Section 2.1).
The average radio emission and 1.4GHz-to-1450$\rm \AA$ ratios 
for each group are calculated using the same method.

The average FIR luminosities (42.5$\mu$m -- 122.5$\mu$m, rest frame) for each
group can be estimated\footnote{A typical redshift
value of z=6 is adopted here} as $\rm 
<L_{FIR}>=2.34\times10^{12}(<f_{250GHz}>/mJy)\,L_{\odot}$, assuming a
dust temperature of 47 K and an emissivity index $\rm \beta$ of 1.6
(Beelen et al. 2006). 
Based on this estimation, the mean
FIR-to-radio ratios for each group can be calculated as (Helou et al. 1985):
\begin{equation}
\rm q\equiv log\left(\frac{\langle L_{FIR}\rangle}{3.75\cdot10^{12}\,L_{\odot}}\right)-log\left(\frac{\langle L_{1.4GHz}\rangle}{L_{\odot}\cdot Hz^{-1}}\right)
\end{equation}
The rest-frame 1.4 GHz radio luminosity densities ($\rm L_{1.4GHz}$) are derived with
the average radio flux densities, assuming a power-law radio spectrum 
of $\rm f_{\nu}\sim\nu^{-0.75}$ (Condon 1992). The q values can be increased by $\sim$0.2 
if we assume a flater radio spectrum of $\rm f_{\nu}\sim\nu^{-0.5}$.
The mean flux densities,
luminosities, and q values are listed in Table 3. We exclude the
three radio loud sources in the first and the third groups, and present the corresponding
results in the second row of each group in Table 3. The ten MAMBO
detected sources in the second group are all radio quiet (according 
to the definition in Kellermann et al. 1989).

\subsection{The mean SED}

The mean 250 GHz flux density is $\rm 1.26\pm0.10\, mJy$ for
the whole sample, and $\rm 2.30\pm0.15\, mJy$ 
for the MAMBO detections. The mean 250 GHz 
flux density of MAMBO non-detections is recovered by 
averaging all the on-source pixel values. A 
$\rm \sim4\sigma$ signal of $\rm 0.52\pm0.13\, mJy$ is 
obtained for this group, which is about one quarter of the value of 
the typical MAMBO detections. We ran the stacking with 
the non-detection group several times, omitting one source 
each stack, to see if this $\rm 4\sigma$ signal is due to a
single source that is just below the detectability.
The test shows stacking flux densities in the range 
of 0.47 to 0.56 mJy with rms values of $\sim$0.13 mJy. Thus it 
is unlikely that the average flux density is dominated by one object. 
No significant differences are found in the $\rm \langle f_{250GHz}\rangle$ values when
the radio-loud sources are excluded in each group. 

The average 1.4 GHz observed flux densities are $\rm 27\pm4$ to $\rm 37\pm4\, \mu Jy$ 
for the radio quiet sources in the three groups. We convert these measurements 
to the flux densities at rest frame 5 GHz assuming a power-law radio spectrum 
of $\rm f_{\nu}\sim\nu^{-0.75}$ (Condon 1992). This gives an 
average radio loudness (i.e. average radio to optical emission ratio 
$\rm \langle R\rangle=\langle f_{\nu,5GHz}\rangle/\langle f_{\nu,4400}\rangle$) 
of $\rm \langle R\rangle\sim$0.7, which is consistent with the typical radio loudness values of $\rm 0.1<R<1$ 
found with local(Kellermann et al. 1989) and z$\leq$2 (Cirasuolo et al. 2003) radio quiet quasars.
The derived q parameters (i.e. FIR to radio emission ratios) are all lower or at the 
edge of the range of $\rm q=2.34\pm0.7$ for typical star 
forming galaxies (Yun et al. 2001), but are consistent with 
the values of q$<$2 found in the local Seyfert 1 galaxies 
(Ulvestad \& Ho 2001). This result suggests that, as was found in the 
local universe and at z$\rm \sim$2 to 4, 
AGN power dominates the radio emission in a large fraction of these radio quiet 
quasars, i.e., the radio quiet AGNs are in fact not radio silent 
(Barvainis et al. 2005; Carilli et al. 2001; Petric et al. 2006). 

We plot the mean SED (normalized by the rest frame 1450$\rm \AA$ flux density) 
of the radio quiet sources in each group in Figure 2, 
together with the templates of low-z optical quasars
from Elvis et al. (1994) and Richards et al. (2006).
We also plot the average optical to near infrared 
emission based on the Spitzer photometry of 13 z$\sim$6 
quasars from Jiang et al. (2006), who show that the ratio 
is consistent with the local 
templates. The FIR-to-1450$\rm \AA$ ratios of the three z$\sim$6 quasar groups have a 
range of $\sim$0.6 dex, and no obvious excess is seen between the value of the whole 
sample and the FIR end of the templates. 
However, due the lack of millimeter and submillimeter data, the available templates 
do not give a good constraint of the FIR-to-millimeter SED for the local 
optical quasars. Sanders et al. (1989) studied a sample of optically selected 
Palomar-Green quasars (PG quasars; Schmidt \& Green 1983; Boroson \& Green 1992), 
and their analysis suggest that the typical 100$\rm \,\mu$m to 1$\rm \,mm$ spectral 
index $\rm \alpha$ (($\rm f_{\nu}\sim\nu^{\alpha}$)) is $\rm \geq2$. We extrapolate 
the FIR emission of the templates with this result (an example of $\rm \alpha=2$ is plotted in Figure 2).  
The extrapolation matchs the mean value of the z$\sim$6 MAMBO undetected quasars very well, while 
the average value of the MAMBO detection is $\sim$2.5 times higher.

\subsection{The SEDs of objects with 250 GHz detections}

We plot the individual optical-to-radio spectral energy distribution
(SEDs) for the ten sources that were detected by MAMBO in Figure 3. 
The FIR emission from the two strongest MAMBO detections, J1148+5251 
and J0927+2001, was well measured with SHARC-II and SCUBA at shorter
wavelengths, which show FIR emission bumps in the SEDs with dust 
temperatures of 55 K and 52 K, respectively (Robson et al. 2004; 
Beelen et al. 2006; Wang et al. 2008a). Similar excesses are 
also seen in the SEDs of another five sources: J0033-0125, 
J0840+5624, J1335+3533, J1425+3254, and J2054-0005, where the observed 
250 GHz flux densities are $\rm \gtrsim0.5$ dex higher than those 
expected from the quasar templates fit to the observed 
optical emission. 

We model the FIR emission of these sources with a optically thin
graybody assuming an emissivity index of $\rm \beta=1.6$ (Beelen et
al. 2006). For J0927+2001 and J1148+5251, we directly adopt the
dust temperatures fit in Wang et al. (2008a) and Beelen et al. (2006), respectively.
For the other nine sources, we assume a dust temperature of 47K (Beelen et al. 2006;
Wang et al. 2007). The model FIR SED was extrapolated to the radio band
using the typical FIR to radio correlation defined by star forming
galaxies, i.e. q=2.34 (Yun et al. 2001). A radio spectral index of
-0.75 is adopted here (Condon 1992). Three out of the seven FIR excess 
sources have been detected at 1.4GHz with deep VLA 
observations, and the observed flux densities 
are all above the average for their 250 GHz flux and q=2.34, 
but the values are within the range in which star
forming galaxies lie.

\section{Correlations between FIR emission and the AGN}

To investigate the origin of the FIR emission, 
we compare the correlation between the 
FIR luminosities and the central AGN bolometric luminosities of these 
z$\sim$6 quasars to that of two samples of local quasars: 
a sample of forty-five local (z$<$0.5) PG quasars  
from Hao et al. (2005), a sample of thirty-one z$<$0.35 type I AGNs hosted in ULIRGs
with IR luminosities $\rm L(8-1000\mu m)>10^{12}L_{\odot}$
(IR quasars; Zheng et al. 2002; Hao et al. 2005). 

Hao et al. (2005) reported that the IR quasar sample showed stronger FIR emission and 
a shallower slope in the 60 $\mu$m-optical luminosity relationship 
when compared to the PG quasars (see also Hao et al. 2007). 
They attributed this result to the starburst-dominated FIR dust heating 
in the host galaxies of these IR quasars. 
We re-fit these FIR-AGN relationships using the FIR and AGN 
bolometric ($\rm L_{bol}$) luminosities of the two samples. The bolometric
luminosities are taken from Hao et al. (2005), which were derived from the 
extinction-corrected quasar optical emission (see also Zheng et al. 2002). 
The FIR luminosities of the IR quasars are taken from Zheng et al. (2002), 
while for the PG quasar sample we adopt the FIR luminosities and upper limits presented
in Haas et al. (2003).  All these data are corrected to our cosmology.
We adopt the Expectation-Maximization 
method in the IRAF STSDAS package (Isobe et al. 1986) which can perform linear 
regression with censored data. The results are: \\
\begin{equation}
\rm IR\,quasars:\,log L_{FIR}=(0.35\pm0.08)log L_{bol}+(7.9\pm1.0)
\end{equation}
\begin{equation}
\rm PG\,quasars:\,log L_{FIR}=(0.66\pm0.08)log L_{bol}+(2.8\pm1.0)
\end{equation}

Figure 4 compares the z$\sim$6 quasars to these two local 
quasar samples in the $\rm L_{FIR}-L_{bol}$ plane.
All the millimeter detected sources, and the average of the 
non-detections are presented. For millimeter detections, 
$\rm L_{FIR}$ was integrated 
using the modeled FIR SEDs described in Section 3.2 
(see also Wang et al. 2007), and for non-detections we 
adopt the estimation of average FIR luminosity in 
Section 3. The AGN bolometric luminosities of these z$\sim$6 
quasars are estimated based on optical and near-infrared 
observations. Jiang et al. (2006) presented $\rm L_{bol}$ for the  
Spitzer observed sources. We adopted their results 
(corrected to the cosmology we adopted in this paper) and estimated
$\rm L_{bol}$ from the optical B-band luminosity for the other sources,
$\rm L_{bol}=10.4L_{B}$ (Richards et al. 2006). The B-band 
luminosities are estimated with the rest frame 1450$\rm \AA$ 
magnitudes assuming a power-law spectrum of $\rm f_{\nu}\sim\nu^{-0.5}$. 
We also convert $\rm L_{bol}$ to
black hole accretion rate $\rm \dot M$ using\\
\begin{equation}
\rm L_{bol}= \eta \dot M c^{2},
\end{equation}
and an assumed efficiency $\eta=0.1$. The derived parameters
for the ten sources are listed in Table 4.

Nine of the ten MAMBO-detected z$\sim$6 quasars stand
close to the relationship derived from local IR quasars in the $\rm L_{FIR}-L_{bol}$ plot (Figure 4), 
while the other source J0818+0722 has a lower FIR luminosity, 
falling between the relationships of the IR and PG quasars. 
This may suggest that the strong millimeter detections 
at z$\sim$6 are the high-mass counterparts of the local 
IR selected quasars, and a similar starburst dust-heating 
mechanism dominates their FIR emission. 
However, these z$\sim$6 quasars are much more luminous in the optical than the 
two local quasars and the $\rm L_{FIR}-L_{bol}$ trends 
derived with the two local samples are getting mixed at 
high luminosity end. Thus extending the millimeter study to 
the optically fainter quasar population at z$\sim$6 will help to determine 
the $\rm L_{FIR}-L_{bol}$ correlation for the millimeter bright z$\sim$6 quasars  
and make a better comparison at low and high redshifts. 
On the other hand, the average value of the MAMBO non-detections is consistent 
with the trend defined by the local PG quasars.

\section{A link between millimeter detection and the strength of $\rm Ly\alpha$ emission}

Omont et al. (1996b) studied a sample of z$\sim$4 optically bright 
quasars and found that the millimeter detected sources tend to have weaker 
UV emission lines compared to the average line strength of the millimeter non-detections.
Bertoldi et al. (2003a) found a similar trend with MAMBO observations
of five z$\sim$6 quasars. In this section, we investigate this
effect with the $\rm Ly\alpha$ emission of full sample of z$\sim$6 quasars observed with MAMBO. The rest-frame $\rm Ly\alpha$
equivalent widths ($\rm EW_{Ly\alpha}$) of five quasars in the
SDSS deep strip were presented in Jiang et al. (2007a). 
The UV spectra of another 20 sources are available in the papers of  
Fan et al. (2000, 2001a, 2003, 2004, 2006a, 2008, in prep). We estimate
$\rm EW_{Ly\alpha}$ for 18 of them. The spectra have been 
corrected for Galactic extinction, adopting the extinction curve 
presented in Savage \& Mathis (1979). We then fit the spectra using a  
method similar to that used in Jiang et al. (2007a), i.e. a power-law 
($\rm f_{\nu}\sim \nu^{-0.5}$) for the UV
continuum and gaussians for the $\rm Ly\alpha$ and $\rm NV$
line emission. The sources J1044-0125 and J1048+4637 
have broad absorption line features and their equivalent 
widths are therefore unavailable. 

We plot the histograms of $\rm EW_{Ly\alpha}$ for MAMBO detections
(solid line) and non-detections (dotted line) in Figure 5. 
Most of the MAMBO detections distribute in the region 
of $\rm log\,EW_{Ly\alpha}(\AA)<1.5$ with a median 
value of 24 $\rm \AA$, while most non-detections 
have $\rm log\,EW_{Ly\alpha}(\AA)$ $\rm >1.5$ with a 
median value of 68 $\rm \AA$ which is comparable 
to the average value of $\sim$70 $\rm \AA$ from the large optical 
quasar samples at lower redshifts (Schneider et al. 1991; Fan et al. 2001b). 
However, there is also one MAMBO 
non-detected source, J1621+5155, which shows very weak $\rm Ly\alpha$
emission with $\rm EW_{Ly\alpha} < 5\AA$.

The physical origin of this curious trend is not clear. 
According to the spectra study with samples of z$\sim$4 quasars presented 
in Omont et al. (1996b), similar weaker emission is also seen in other UV 
emission lines, such as the Si \textsc{\small IV}/O \textsc{\small IV]} and C \textsc{\small IV} lines. 
Though the UV line emission is weak in the MAMBO detected sources,
there is no evidence of strong dust reddening in the UV continuum.
We can speculate that intrinsically different physical conditions associated with the broad
line emission cloud, or a special dust obscuration geometry that affects 
only the broad line region, may explain this effect. Further tests 
with other strong broad lines, such as the C \textsc{\small IV}, and Mg \textsc{\small II} lines in 
the UV, the H$\alpha$, and H$\beta$ lines in the 
optical, and the Paschen lines in near-infrared, are required to
give a better understand of this trend. In particular, the  
trend may disappear when tested with the near-infrared lines if it is 
due to dust absorption. 

\section{Discussion}

In this paper, we analyze the FIR properties of the current quasar
sample at z$\sim$6. These quasars are mainly optically selected, and
represent the luminous end of the quasar population at this early epoch. 
The X-ray to near-infrared SEDs of these sources are  
dominated by powerful AGN, just as in  
optical quasars at lower redshift (Jiang et al. 2006). 
We divide the sample into MAMBO 250 GHz detections of 10 sources 
and non-detections of 23 sources. Their FIR emission was compared to 
that of the low-z quasars, and with the results listed above we will further discuss 
the FIR dust heating and evolution stage of these z$\sim$6 quasars in this section. 

\subsection{The average FIR emission of the 250 GHz non-detections}

The average FIR emission of the MAMBO non-detections is 
recovered at $\rm 4\sigma$ by stacking the 250 GHz 
measurements of each source, and the mean SED is similar 
to the low-z quasar templates of Elvis et al. (1994) 
and Richards et al. (2006). This suggests that the emission 
from the outer part of AGN-heated dust torus may dominates 
the FIR SED in these sources, as was found in the low-z 
optically bright quasars (Sanders et al. 1989; Haas et al. 2003).
However, we cannot rule out contributions from star formation in the
quasar hosts. Further surveys of PAH and other star forming features 
will address the average
contribution from star formation to FIR dust heating in these objects. 
The average FIR luminosity (42.5$\mu$m -- 122.5$\mu$m, rest frame) derived from the stacking flux 
density is 1.2$\rm \times10^{12}\,L_{\odot}$. 
If star formation is not the dominating dust heating 
source (i.e. contributes $<$50\%), the average star formation rate in the quasar 
hosts\footnote{The star formation rate is estimated with
the empirical relationship from Kennicutt
(1998), assuming a standard Salpeter initial mass 
function, i.e. $\rm SFR\sim 4.5L_{IR}\,M_{\odot}yr^{-1}$,
where $\rm L_{IR}$ is the infrared
luminosity (8-1000$\,\mu$m) in unit
of $\rm 10^{44}\,erg\,s^{-1}$, and
is $\rm\sim 1.5L_{FIR}$ for warm dust emission.} 
should be less than $\rm 200\,M_{\odot}\,yr^{-1}$.

\subsection{The FIR emission of the strong 250 GHz detections}

Excess FIR emission is seen in seven of the MAMBO detected sources (see Figure 3).
Their 250 GHz flux densities are much stronger ($\rm \gtrsim0.5$ dex)
than the values expected from the quasar templates when normalized  
in the rest-frame UV. These sources
show FIR-to-radio luminosity ratios within the range defined by typical
star forming galaxies (Section 3.2, see also Carilli et al. 2004;
Wang et al. 2007). The relation between FIR and AGN bolometric luminosity for 
these sources follows the trend defined by a sample of low-redshift IR luminous
quasars which are hosted by ULIRGs (Section 4). The Spitzer
near-infrared photometry by Jiang et al. (2006) detected the 
AGN-heated hot dust ($\sim$1000 K) in some of these bright millimeter 
sources; the optical-to-near infrared SEDs of these sources 
match the low-z templates very well. The strong FIR emission bump in
these sources indicates an additional warm dust component, with 
a dust temperature of 50 to 60 K. Brightness temperature 
arguments imply that the warm dust emission region must be 
extended over a few kpc (Wang et al. 2008a). Thus either a 
significant fraction of the dust heating 
is due to star formation in the host galaxy, or the  
AGN must heat dust over large scales. 

Intense star formation activity is suggested by other observations of 
these strong millimeter quasars. Molecular CO emission is 
detected in the three strongest z$\sim$6 millimeter sources:  
J1148+5251 (Bertoldi et al. 2003b; Walter et al. 2003, 2004),  
J0927+2001 (Carilli et al. 2007), and J0840+5624 (Wang et al. 2008b, in prep).
These CO detections indicate the presence of 
$\rm \gtrsim10^{10}\,M_{\odot}$ of molecular 
gas in the quasar hosts (Walter et al. 2003; Bertoldi et al. 2003b; Carilli et al. 2007), 
which can provide the fuel required for star formation rates 
of up to $\rm >10^{3}\,M_{\odot}\,yr^{-1}$. 
Indeed, it is hard to see how the host galaxies can avoid massive 
star formation with so much molecular gas.
The measurement of several CO transitions in J1148+5251 suggests a
gas density of about $\rm 10^{4.5}\,cm^{-3}$ and a kinetic 
temperature of $\sim$100 K (Bertoldi et al. 2003b), which are
similar to values in starburst environments found in other quasars and
submillimeter galaxies at lower redshifts (Carilli et al. 2002;
Solomon \& Vanden Bout 2005; Riechers et al. 2006). In addition, 
recent high resolution observations of [C {\small II}] in J1148+5251 
(Walter et al. 2008, in prep.; Carilli 2008) show 
an emission region extended over many kpc, i.e., comparable to that of 
the CO (3-2) line. 

Further studies also suggest connections between the CO and FIR 
emission. The FIR-to-CO luminosity ratios 
of J1148+5251 and J0927+2001 are about 500 and 650 
$\rm L_{\odot}$ $\rm (K\,km\,s^{-1}\,pc^{2})^{-1}$, respectively, which are
consistent with the values found in other high-z CO 
emitting galaxies (Solomon \& Vanden Bout 2005).
In Figure 6, we indicate these two CO detected z$\sim$6
quasars (as stars) on the FIR-CO luminosity
correlation plot discussed in Riechers et al. (2006).
The two points fall a little above but within the observed scatters of the FIR-CO luminosity relation 
of the typical star forming galaxies at both low and 
high redshifts. 

All these facts suggest
that massive star formation is an important source of
the FIR dust heating. However, it is difficult to constrain what fraction of the FIR
dust heating is contributed by star formation in these FIR
luminous quasars. According to the
FIR-to-CO luminosity correlation in Figure 6, at least $\rm 30-50\%$
of the FIR emission should originate from star formation, giving a 
corresponding star formation rate of $\rm \gtrsim 1000\,M_{\odot}\,yr^{-1}$.

\subsection{The evolutionary stage of the z$\sim$6 quasars}

Observations from optical to near-infrared of these z$\sim$6 quasars 
indicate that the central SMBHs have masses of $\rm \gtrsim10^{9}\,M_{\odot}$ 
(Willott et al. 2003; Jiang et al. 2006). 
Based on the  
SMBH-bulge mass relationship (eg. Marconi \& Hunt 2003), the final stellar 
bulge mass should be on order of $\rm 10^{12}\,M_{\odot}$ for a mature system. 
The formation of such a massive stellar system may require a 
star formation rate of $\rm \gtrsim10^{3}\,M_{\odot}\,yr^{-1}$
on time scales of a few hundred million years as was suggested 
by some numerical simulations (eg. Li et al. 2007a, b). 
Thus the average star formation rate 
of $\rm <200\, M_{\odot}\,yr^{-1}$ found in the 250 GHz 
non-detected quasars can be either the case that 
the major bulge building has already been finished 
and the SMBH-bulge mass relationship seen at z=0
has already been established, or that the host galaxy 
is has not yet reached the massive star formation phase,  
and the growth of the stellar bulge is  
lagging behind that of the SMBH (see eg. Shields et al. 2006; Ho 2007). 
Deep and high-resolution imaging at near-IR wavelengths may provide a direct measurement of 
the stellar population in these z$\sim$6 quasar hosts.

For the strong millimeter quasars at z$\sim$6, the FIR excess 
suggests a star formation rate of $\rm 
\gtrsim1000\,M_{\odot}\,yr^{-1}$ from the host galaxies. 
It is likely that we are witnessing simultaneous building of the 
stellar bulge and growth of the SMBH in these objects.
The case of  J1148+5251 is the best documented:
Willott et al. (2003) derived an estimate 
of $\rm 3\times10^9\,M_{\odot}$ for the black hole mass, while
Walter et al. (2004) estimated a dynamical mass 
of $\rm \sim5\times10^{10}\,M_{\odot}$. The SMBH-bulge 
mass ratio is thus clearly more than one order of magnitude larger than
the local SMBH-bulge relationship (eg. Marconi \& Hunt 2003), as in
several other high-z CO-detected QSOs (Shields et al. 2006; Ho 2007).
However, the amount of gas detected in CO observations
such as in J1148+5251, a few $\rm 10^{10}\,M_{\odot}$, is far short 
of the value required to reach the bulge mass given by the local 
black hole-bulge relationship.

To fully understand the evolutionary stage of
these strong millimeter quasars at z$\sim$6, further
observations are required. First, CO and [C
{\small II}] searches and excitation studies should be extended to all the 
z$\sim$6 quasars with strong FIR excesses to
determine the host galaxy ISM properties. Second, high
resolution ($\rm \lesssim0.3''$) CO, [C {\small II}] and dust mapping should be made
of all the CO detections. If the FIR-emitting
dust and molecular gas are distributed similarly, the FIR emission 
is likely to be due to a starburst, 
while more compact FIR emission will argue for AGN dust
heating. The resolved CO emission would also provide an estimate of the dynamical mass
of the quasar hosts on kpc scales. This is a direct, and perhaps 
the only, way to test the black hole-bulge relationship at 
the highest redshifts. These 
observations can be done with sensitive submillimeter,
millimeter and radio telescopes such as the PdBI, CARMA, and SMA, and the
coming ALMA and EVLA.

\section{Conclusion}

We study the SEDs of 33 quasars at z$\sim$6 from FIR to radio wavelengths. 
We conclude that, when averaged with the whole sample and the 
250 GHz undetected sources, no significant difference is seen between the mean 
FIR-to-radio SEDs and the templates of low-redshift optically selected quasars.
In particular, we extrapolate the FIR emission of the templates with the typical 
quasar FIR-to-millimeter spectrum (i.e. $\rm f\sim\nu^\alpha,\,\alpha\geq2$) 
from Sanders et al. (1989). This extrapolation is consistent with the average FIR 
emission of the 250 GHz undetected z$\sim$6 quasars very well (see Section 3.1 and Figure 2). 
This fact suggests that these sources have a similar AGN dominated dust
heating mechanism. The average star formation rate from the
host galaxies of the quasars undetected by MAMBO is estimated to be less than a few hundred
$\rm M_{\odot}\,yr^{-1}$. 

We detect a strong FIR excess in seven of
the ten z$\sim$6 quasars detected in the millimeter band. These FIR luminous sources are likely
to be the high-mass counterparts of local IR luminous quasars which
are in transition between the starburst phase and the mature 
quasar phase. Star formation at rates of $\rm >10^{3}\,M_{\odot}\,yr^{-1}$ 
may be the dominant mechanism to heat the dust that gives rise to the FIR emission.
The millimeter detected sources also show weaker Lya emission
compared to that of the non-detections, but the origin of this trend is not clear yet.
Further observations of these FIR luminous quasars, especially spatially resolved 
line and dust studies, are required to further 
address questions of star formation and FIR dust heating in these
extreme objects.

The current sample of z$\sim$6
quasars is optically selected. 
There may be other quasar populations at z$\gtrsim$6 
which are still buried in their starburst environment and are obscured in
the optical. Sensitive IR and millimeter facilities,
such as Spitzer, ALMA, and Herschel, may discover such objects and provide a more 
complete view of galaxy and SMBH evolution in the early 
universe.

\acknowledgments 
We acknowledge support from the Max-Planck Society and the Alexander von Humboldt
Foundation through the Max-Planck-Forschungspreis 2005. 
Michael, A. Strauss acknowledges support of the National Science
Foundation grant AST-0707266. 
X. Fan acknowledge support from a David and Lucile Packard
Fellow in Science and Engineering. We thank James J. Condon for comments and suggestions. 
IRAM is funded by the Centre National de la Recherche Scientifique (France), 
the Max-Planck Gesellschaft (Germany), and the Instituto Geografico Nacional (Spain).
The National
Radio Astronomy Observatory is a facility of the National Science
Foundation, operated by Associated Universities, Inc.

Funding for the SDSS and SDSS-II has been provided by the Alfred P. 
Sloan Foundation, the Participating Institutions, the National Science 
Foundation, the U.S. Department of Energy, the National Aeronautics and 
Space Administration, the Japanese Monbukagakusho, the Max Planck 
Society, and the Higher Education Funding Council for England. The SDSS 
Web Site is http://www.sdss.org/.
The SDSS is managed by the Astrophysical Research Consortium for the 
Participating Institutions. The Participating Institutions are the 
American Museum of Natural History, Astrophysical Institute Potsdam, 
University of Basel, University of Cambridge, Case Western Reserve 
University, University of Chicago, Drexel University, Fermilab, the 
Institute for Advanced Study, the Japan Participation Group, Johns 
Hopkins University, the Joint Institute for Nuclear Astrophysics, the 
Kavli Institute for Particle Astrophysics and Cosmology, the Korean 
Scientist Group, the Chinese Academy of Sciences (LAMOST), Los Alamos 
National Laboratory, the Max-Planck-Institute for Astronomy (MPIA), the 
Max-Planck-Institute for Astrophysics (MPA), New Mexico State 
University, Ohio State University, University of Pittsburgh, University 
of Portsmouth, Princeton University, the United States Naval 
Observatory, and the University of Washington.

{\it Facilities:} \facility{IRAM:30m (MAMBO)}, \facility{VLA}, \facility{Sloan (SDSS)}


\begin{table}
{\scriptsize \caption{Summary of the new observations \label{tbl-1}}
\begin{tabular}{lcccccc}
\hline \noalign{\smallskip}
\hline \noalign{\smallskip}
Name & redshift &$\rm m_{1450}$ & $\rm M_{1450}$ & $\rm f_{250GHz}$ & $\rm f_{1.4GHz}$ & Reference$^{a}$ \\
     &   &                 &  &   mJy            &  $\rm \mu$Jy & \\
(1) & (2) & (3) & (4) & (5) & (6) & (7) \\
\noalign{\smallskip} \hline \noalign{\smallskip}
J000552.34$-$000655.8 & 5.85&20.23 &-26.47 & 0.36$\pm$0.48 & 40$\pm$130& 1,$\ast$,2\\
J003311.40$-$012524.9 & 6.13&21.78 &-25.00 &{\bf 1.13$\pm$0.36} &-27$\pm$19 & 3,3,$\ast$\\
J020332.39+001229.3 & 5.86&20.97 &-25.73 & 1.52$\pm$0.67 & {\bf 195$\pm$22} & 4,$\ast$,$\ast$\\
J030331.40$-$001912.9 & 6.07&21.33 &-25.43 & 0.23$\pm$0.51 & -85$\pm$62 & 4,$\ast$,$\ast$\\
J035349.76+010405.4 & 6.05&20.21 &-26.55 & 1.20$\pm$0.46 &17$\pm$19 & 4,$\ast$,$\ast$\\
J081827.40+172251.8 & 6.00&19.34 &-27.40 & {\bf 1.19$\pm$0.38} &{\bf 123$\pm$12} & 5,$\ast$,2\\
J084119.52+290504.5 & 5.96&19.61 &-27.12 & 1.00$\pm$0.43 &43$\pm$27 & 8,$\ast$,$\ast$\\
J084229.23+121848.2 & 6.08&19.58 &-27.18 & 0.11$\pm$0.55 &-8$\pm$19 & 9,$\ast$,$\ast$\\
J092721.82+200123.7 & 5.77&19.87 &-26.81 & {\bf 4.98$\pm$0.75} &{\bf 50$\pm$11} & 5,2,$\ast$\\
J104433.04$-$012502.2 & 5.78&19.21 &-27.46 & {\bf 1.82$\pm$0.43} & -15$\pm$24 & 10,$\ast$,7\\
J142516.30+325409.0 & 5.85&20.62 &-26.08 & {\bf 2.27$\pm$0.51} &20$\pm$20 & 14,$\ast$,$\ast$\\
J142738.59+331242.0 & 6.12&20.33 &-26.44 & 0.39$\pm$0.66 &{\bf 1730$\pm$131} & 15,$\ast$,15\\
J160253.98+422824.9 & 6.07&19.86 &-26.90 & 1.41$\pm$0.54 &{\bf 60$\pm$15} & 1,$\ast$,2\\
J162100.70+515544.8 & 5.71&19.89 &-26.77 & 0.30$\pm$0.55 &-12$\pm$21 & 9,$\ast$,$\ast$\\
J163033.90+401209.6 & 6.05&20.64 &-26.12 & 0.80$\pm$0.60 &14$\pm$15 & 11,12,$\ast$\\
J164121.64+375520.5 & 6.04&21.30 &-25.45 & 0.08$\pm$0.46 &-30$\pm$32  & 3,3,$\ast$\\
J205406.42$-$000514.8 & 6.07&20.67 &-26.09 & {\bf 2.38$\pm$0.53} &17$\pm$23 & 4,$\ast$,$\ast$\\
J231546.36$-$002357.5 & 6.12&21.31 &-25.46 & 0.28$\pm$0.60 &31$\pm$16 & 4,$\ast$,$\ast$\\
J232908.28$-$030158.8 & 6.43&21.65 &-25.20 & 0.01$\pm$0.50 &14$\pm$22 & 3,3,$\ast$\\
\noalign{\smallskip} \hline
\end{tabular}\\
}
{\scriptsize Note -- The detections are marked as boldface.\\
$^{a}$The three references are for the optical, 250 GHz, 
and 1.4 GHz data, respectively. The asterisks denote new data reported in this paper.\\
References for Table 1 and 2 -- (1) Fan et al. 2004; (2) Wang et al. 2007; (3) Willott et al. 2007;
(4) Jiang et al. 2008; (5) Fan et al. 2006a;
(6) Fan et al. 2001a; (7) Petric et al. 2003; (8) Goto 2006; (9) Fan et al.
2008, in prep. (10) Fan et al. 2000; (11) Fan et al 2003; (12) Bertoldi et al.
2003a; (13) Carilli et al. 2004; (14) Cool et al. 2006; (15) McGreer et al. 2006. 
}
\end{table}

\begin{table}
{\scriptsize \caption{Summary of previous observations \label{tbl-2}}
\begin{tabular}{lcccccc}
\hline \noalign{\smallskip}
\hline \noalign{\smallskip}
Name & redshift &$\rm m_{1450}$ & $\rm M_{1450}$ & $\rm f_{250GHz}$ & $\rm f_{1.4GHz}$ & Reference$^{a}$ \\
     &   &                 &  &   mJy            &  $\rm \mu$Jy & \\
(1) & (2) & (3) & (4) & (5) & (6) & (7) \\
\noalign{\smallskip} \hline \noalign{\smallskip}
J000239.39+255034.8 & 5.80&19.02 &-27.67 & 0.20$\pm$0.88 & {\bf 89$\pm$14} & 1,2,2 \\
J083643.85+005453.3 & 5.81&18.81 &-27.88 &-0.39$\pm$0.96 &{\bf 1740$\pm$40} & 6,7,7\\
J084035.09+562419.9 & 5.85&20.04 &-26.66 & {\bf 3.20$\pm$0.64} &12$\pm$9 & 5,2,2\\
J103027.10+052455.0 & 6.31&19.66 &-27.16 &-1.15$\pm$1.13 & -3$\pm$20 & 6,2,2\\
J104845.05+463718.3 & 6.20&19.25 &-27.55 & {\bf 3.00$\pm$0.40} &7$\pm$13 & 11,12,2\\
J113717.73+354956.9 & 6.01&19.63 &-27.12 & 0.10$\pm$1.13 &9$\pm$17 & 5,2,2\\
J114816.64+525150.2 & 6.42&19.03 &-27.82 & {\bf 5.00$\pm$0.60} &{\bf 55$\pm$12} & 11,12,13\\
J125051.93+313021.9 & 6.13&19.64 &-27.14 & 0.07$\pm$0.90 &37$\pm$21 & 5,2,2\\
J130608.26+035626.3 & 6.02&19.55 &-27.19 &-1.05$\pm$1.04 &14$\pm$21 & 6,7,7\\
J133550.81+353315.8 & 5.95&19.89 &-26.84 & {\bf 2.34$\pm$0.50} &{\bf 35$\pm$10} & 5,2,2\\
J141111.29+121737.4 & 5.93&19.97 &-26.75 & 1.00$\pm$0.62 &{\bf 61$\pm$16} & 1,2,2\\
J143611.74+500706.9 & 5.83&20.16 &-26.54 &-0.21$\pm$1.14 & 6$\pm$16 & 5,2,2\\
J150941.78$-$174926.8 & 6.12&19.82 &-26.95 & 1.00$\pm$0.46 &-- & 3,3,-\\
J162331.81+311200.5 & 6.25&20.13 &-26.67 & 0.17$\pm$0.80 &24$\pm$31 & 1,2,2\\
\noalign{\smallskip} \hline
\end{tabular}\\
}
{\scriptsize Note -- The detections are marked as boldface.\\
$^{a}$The three references are for the optical, 250 GHz,
and 1.4 GHz data, respectively, and the corresponding literatures are listed at the end of Table 1.
}
\end{table}

\begin{table}
{\scriptsize \caption{Mean FIR and radio emission of the z$\sim$6 quasars \label{tb3}}
\begin{tabular}{lccccccc}
\hline \noalign{\smallskip}
\hline \noalign{\smallskip}
Group & $\rm Number^{a}$ & $\rm <f_{250GHz}>$ & $\rm <L_{FIR}>$& $\rm Number^{b}$ & $\rm <f_{1.4GHz}>$&$\rm <L_{1.4GHz}>^{c}$ & q \\
    &     & mJy & $\rm 10^{12}L_{\odot}$    &     & $\rm \mu$Jy &$\rm L_{\odot}\,Hz^{-1}$  & \\
(1) & (2) & (3) & (4) & (5) & (6) & (7) & (8)\\
\noalign{\smallskip} \hline \noalign{\smallskip}
Whole sample& 33 & 1.26$\pm$0.10 & 2.9$\pm$0.2 & 32 & 46$\pm$3&0.031$\pm$0.002 &1.41$\pm$0.04 \\ 
(radio quiet)& 30 & 1.29$\pm$0.10 &3.0$\pm$0.2  & 29 & 32$\pm$3 & 0.022$\pm$0.002 &1.57$\pm$0.05 \\
\noalign{\smallskip} \hline \noalign{\smallskip}
MAMBO detections& 10&2.73$\pm$0.06& 6.4$\pm$0.1 & 10 & 37$\pm$4 &0.024$\pm$0.003 & 1.84$\pm$0.05\\
\noalign{\smallskip} \hline \noalign{\smallskip}
Non-detections&23& 0.52$\pm$0.13& 1.2$\pm$0.3 & 22 & 56$\pm$4 &0.038$\pm$0.003 & 0.94$\pm$0.11\\
(radio quiet) &20& 0.51$\pm$0.13& 1.2$\pm$0.3 & 19 & 27$\pm$4 &0.018$\pm$0.003 & 1.24$\pm$0.13\\
\noalign{\smallskip} \hline
\end{tabular}\\
}
{\scriptsize $^{a}$ Number of sources observed at 250 GHz.\\
$^{b}$ Number of sources observed at 1.4 GHz.\\
$^{c}$ A radio spectral index of -0.75 (Condon 1992) is adopted here to calculate the rest frame 1.4 GHz lunimosity.
}
\end{table}

\begin{table}
{\scriptsize \caption{Luminosities of the 250 GHz detected z$\sim$6 quasars \label{tb4}}
\begin{tabular}{cccc}
\hline \noalign{\smallskip}
\hline \noalign{\smallskip}
Name &  $\rm \frac{L_{bol}}{10^{13}L_{\odot}}$&$\rm \frac{\dot M}{M_{\odot}\,yr^{-1}}$ & $\rm \frac{L_{FIR}}{10^{12}L_{\odot}}$\\
(1) & (2) & (3) & (4) \\
\noalign{\smallskip} \hline \noalign{\smallskip}
J003311.40-012524.9   &  1.4&9.4   &   2.6$\pm$0.8\\
J081827.40+172251.8   & 12.8&86.3  &   2.8$\pm$0.9\\
J084035.09+562419.9   &  6.5&43.6  &   7.6$\pm$1.5\\
J092721.82+200123.7   &  7.4&50.0  &  12.2$\pm$2.8\\
J104433.04-012502.2   & 13.5&91.1  &   5.3$\pm$0.8\\
J104845.05+463718.3   & 14.6&98.4  &   6.9$\pm$0.9\\
J114816.64+525150.2   & 18.8&126.8  &  13.4$\pm$3.4\\
J133550.81+353315.8   &  7.6&51.4  &   5.5$\pm$1.2\\
J142516.30+325409.0   &  3.8&25.6  &   5.4$\pm$1.2\\
J205406.42-000514.8   &  3.8&25.8  &   6.8$\pm$1.1\\
\noalign{\smallskip} \hline
\end{tabular}\\
}
{\scriptsize Note -- $\rm {\dot M}$ is the black hole accretion rate derived with equation (7) in Section 4.\\
}
\end{table}

\clearpage

\begin{figure}
\plotone{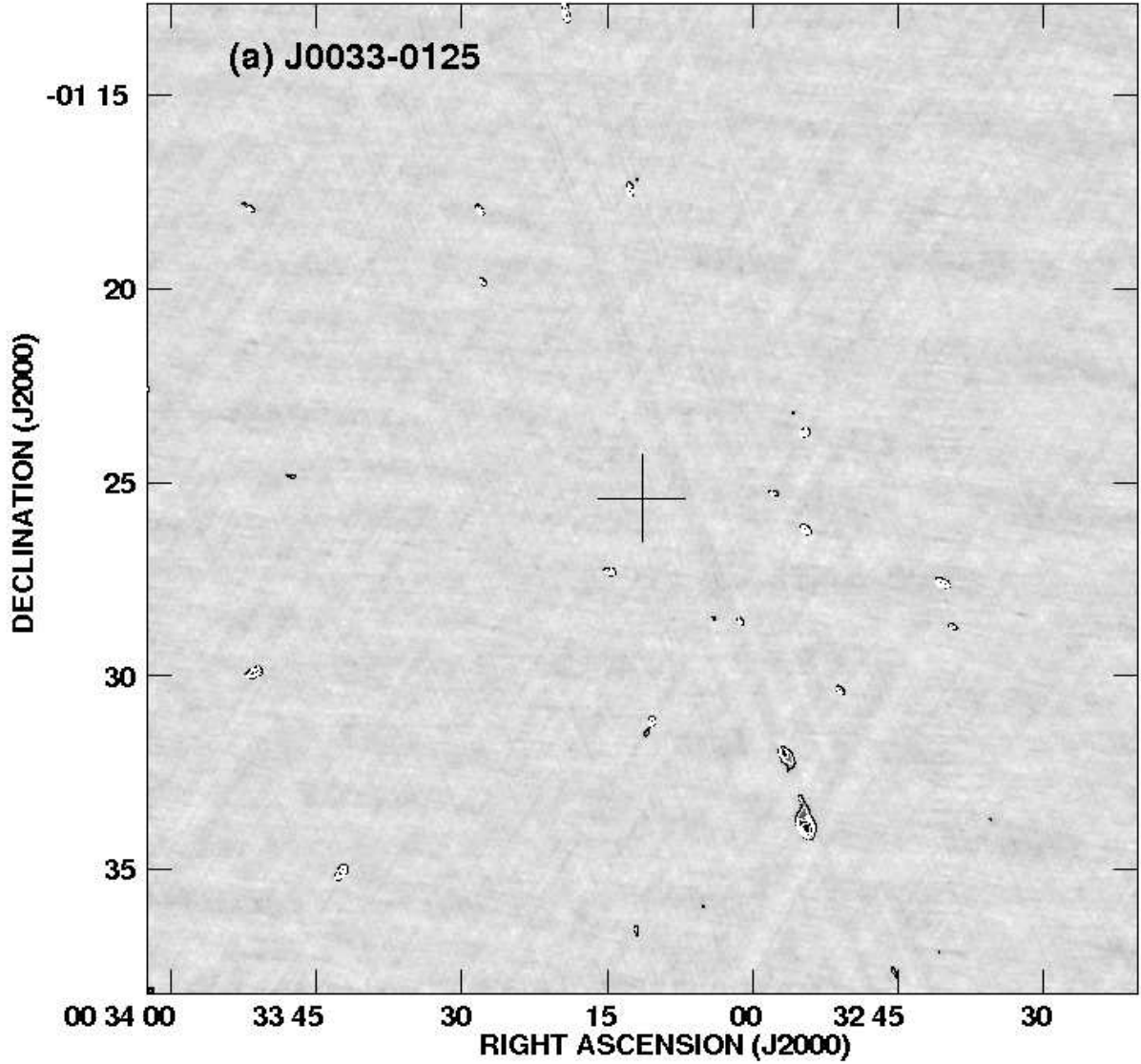}\\
\epsscale{1.0}
\caption{Radio 1.4 GHz field images ($\rm 25'.6\times25'.6$) of J0033-0125 (a) at $\rm 6''.5\times5 ''$ 
resolution (FWHM) and J1641+3755 (b) at $\rm 6''\times4''.5$ resolution. 
Contour levels are a 
geometric progression by a factor of 2, starting at 0.120 mJy$\rm\, beam^{-1}$,
and the crosses denote the optical quasar positions.}
\end{figure}
\begin{figure}
\figurenum{1}
\plotone{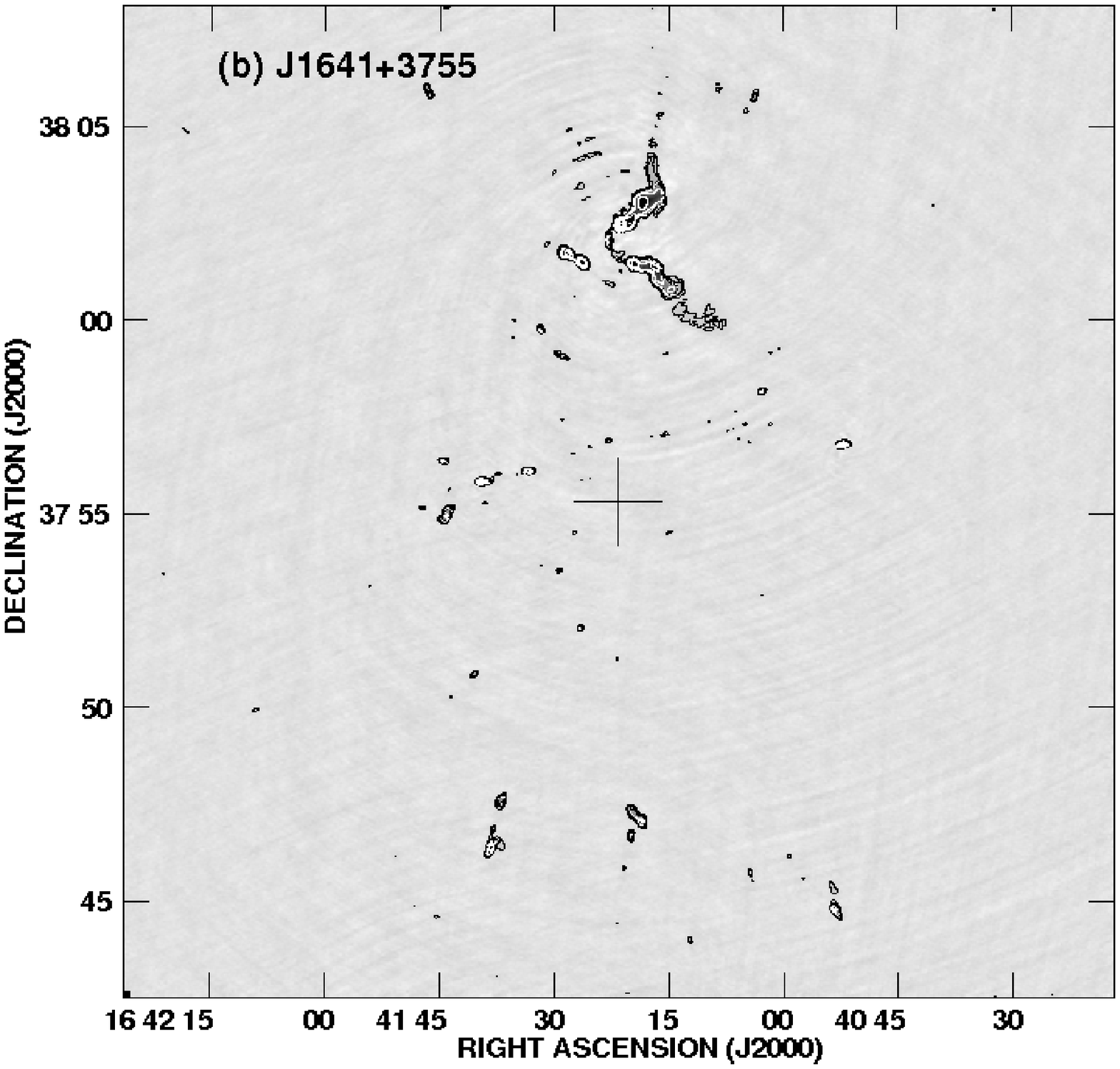}\\
\epsscale{1.0}
\caption{Continued}
\end{figure}

\begin{figure}
\plotone{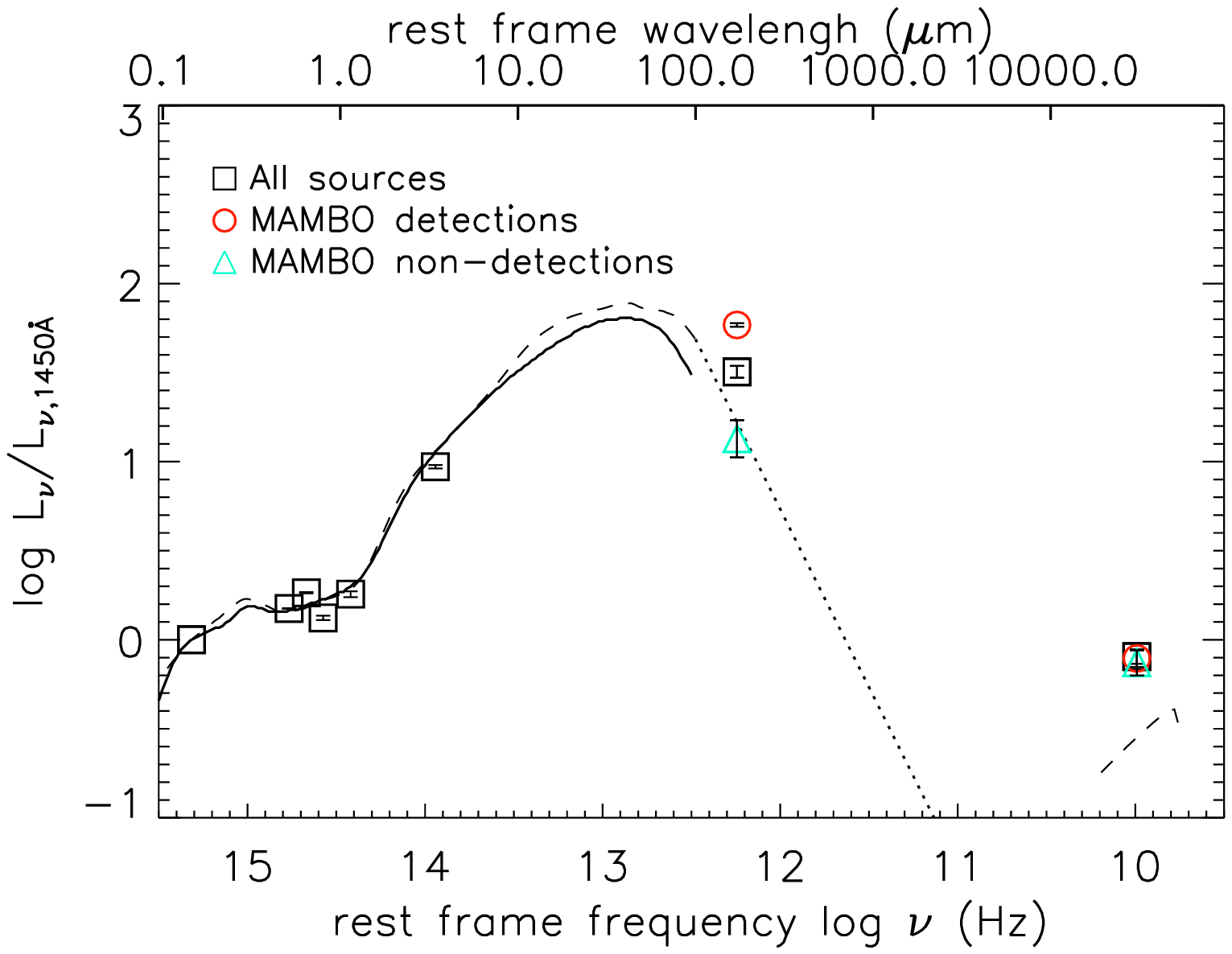}\\
\epsscale{1.0} \caption{Mean FIR and radio emission (normalized by the rest frame $\rm 1450\AA$ luminosity density) of the
radio quiet z$\sim$6 quasars. Black
squares, red circles, and green triangles represent
the average values of the whole z$\sim$6 sample, MAMBO 250 GHz
detections, and non-detections, respectively. 
We also plot the average near-infrared emission (black squares) 
derived from the Spitzer photometry of 13 sources from Jiang et al. (2006). 
The dashed and solid lines are the local quasar templates from Elvis et 
al. (1994) and Richards et al. (2006), respectively. 
The dotted line denotes an extrapolation of the template FIR emission (Elvis et al. 1994) with a power 
law of $\rm L_{\nu}\sim \nu^{2}$  }
\end{figure}

\begin{figure}
\plottwo{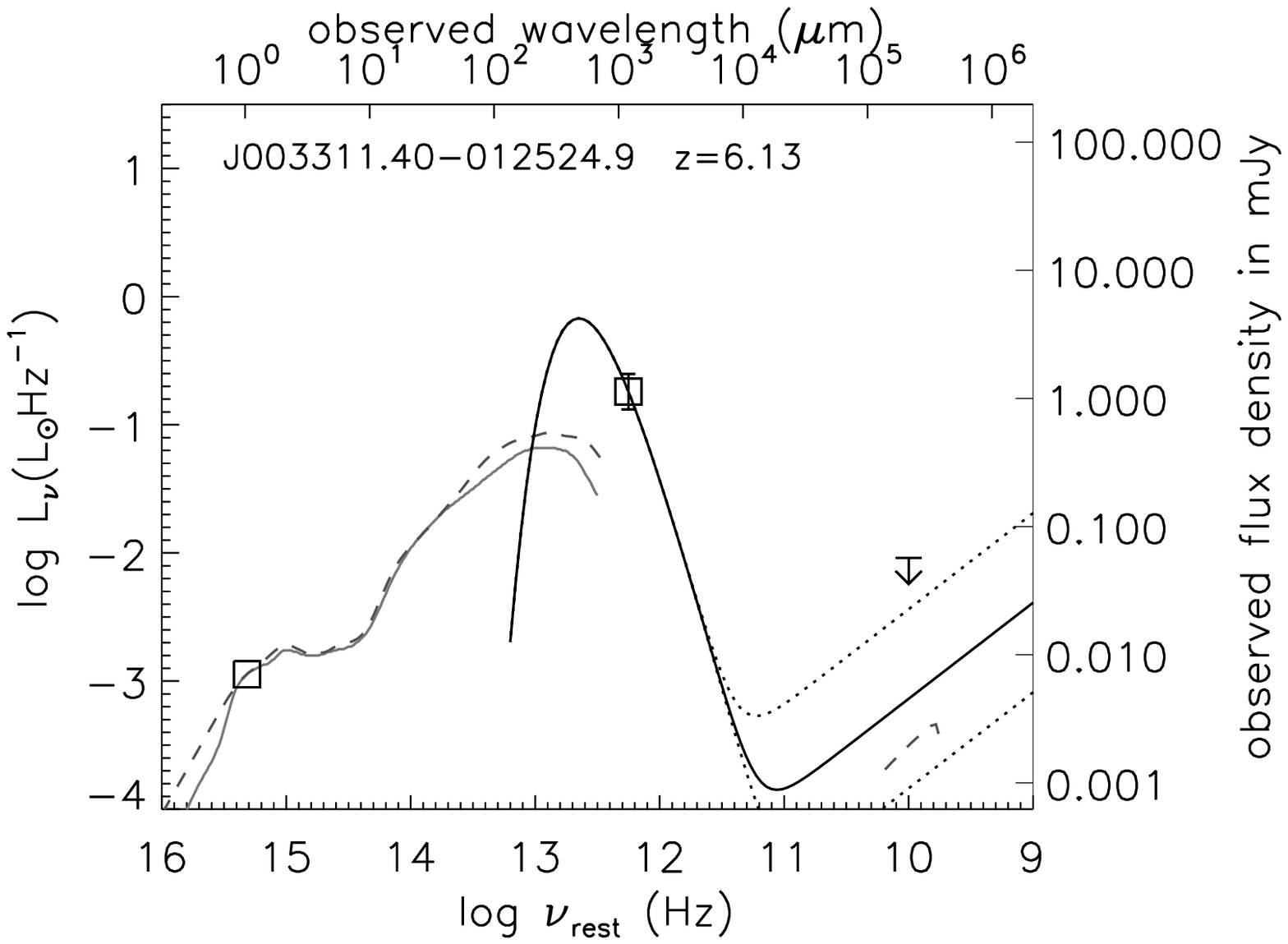}{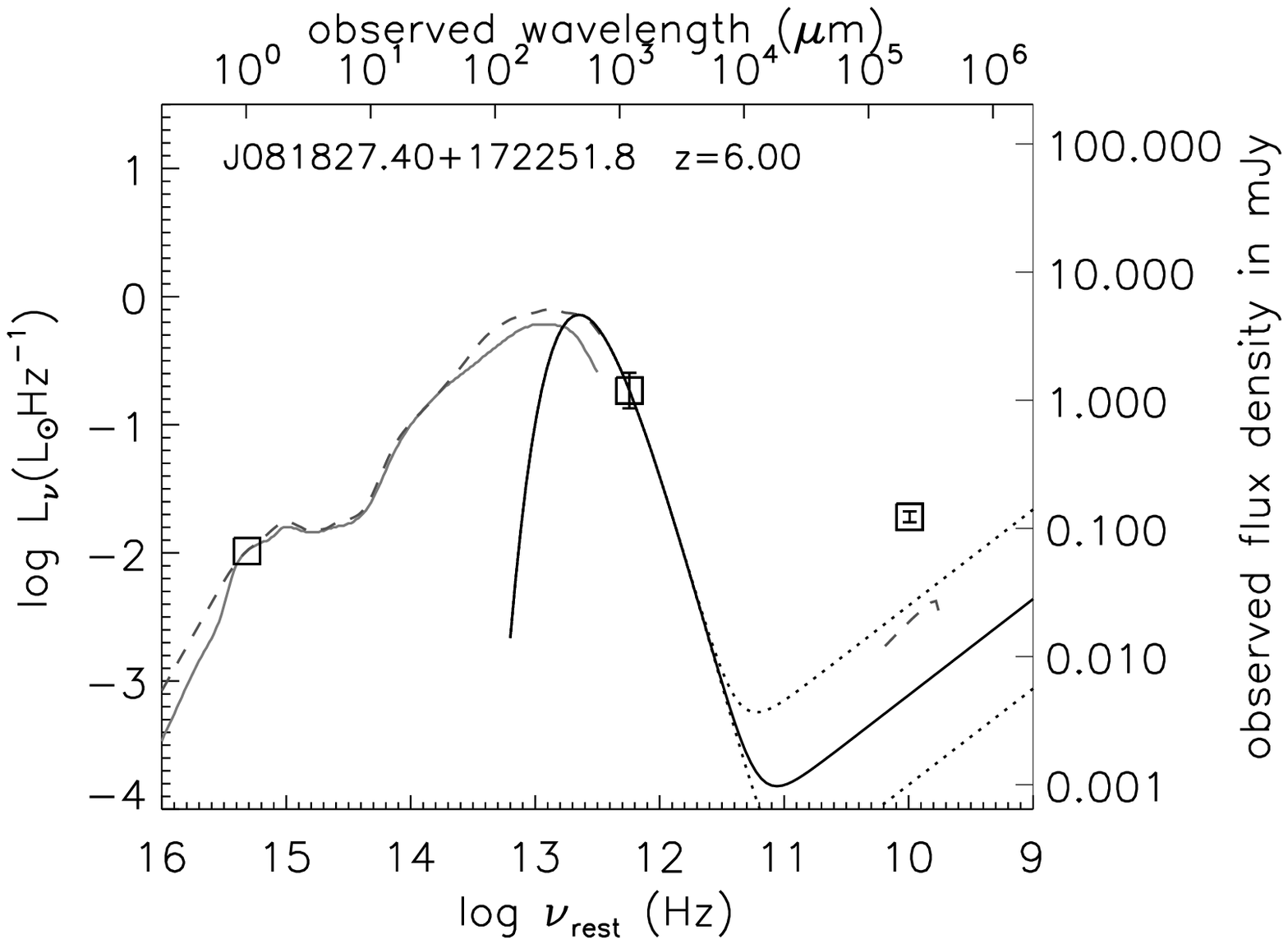}\\
\plottwo{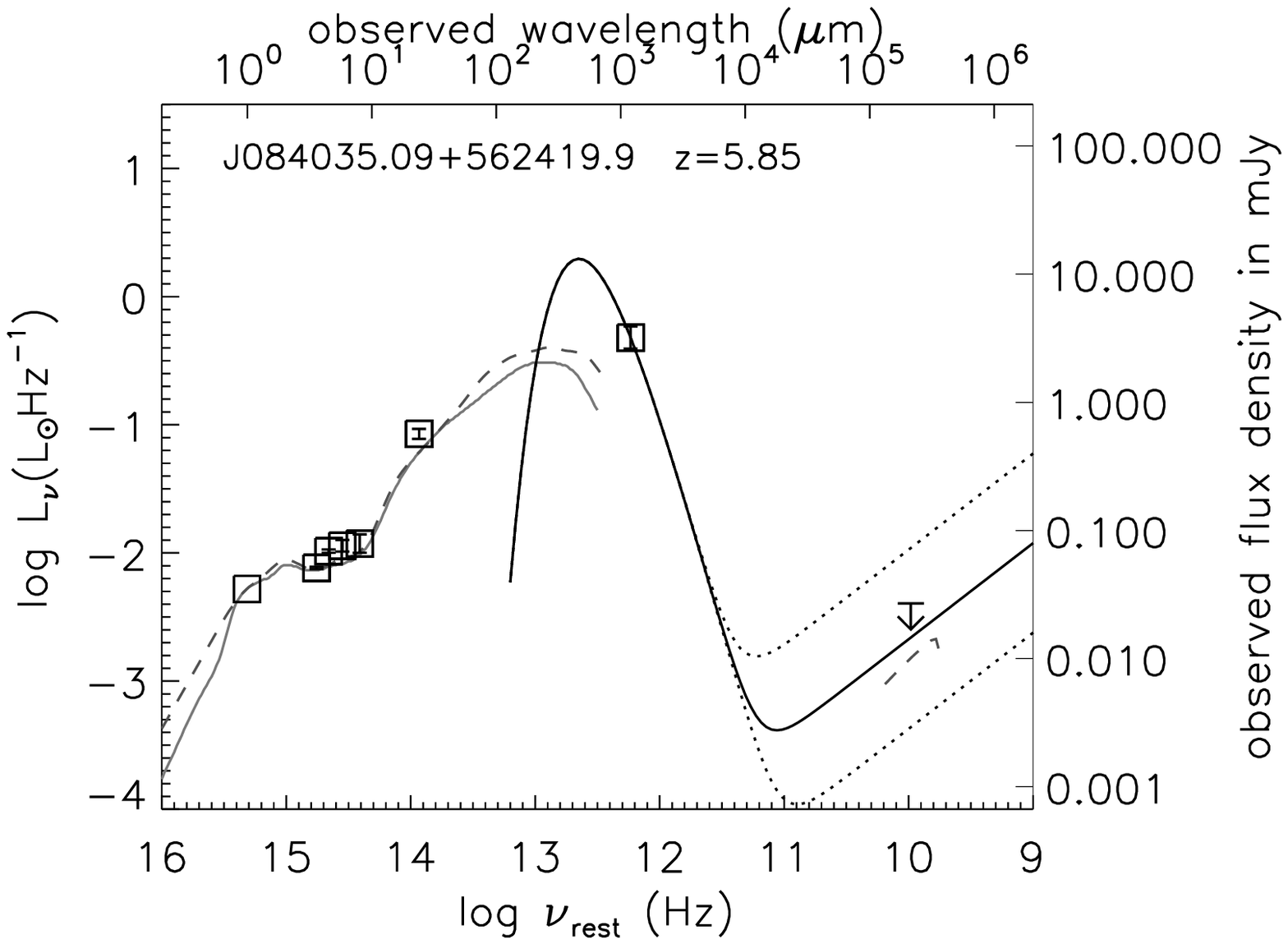}{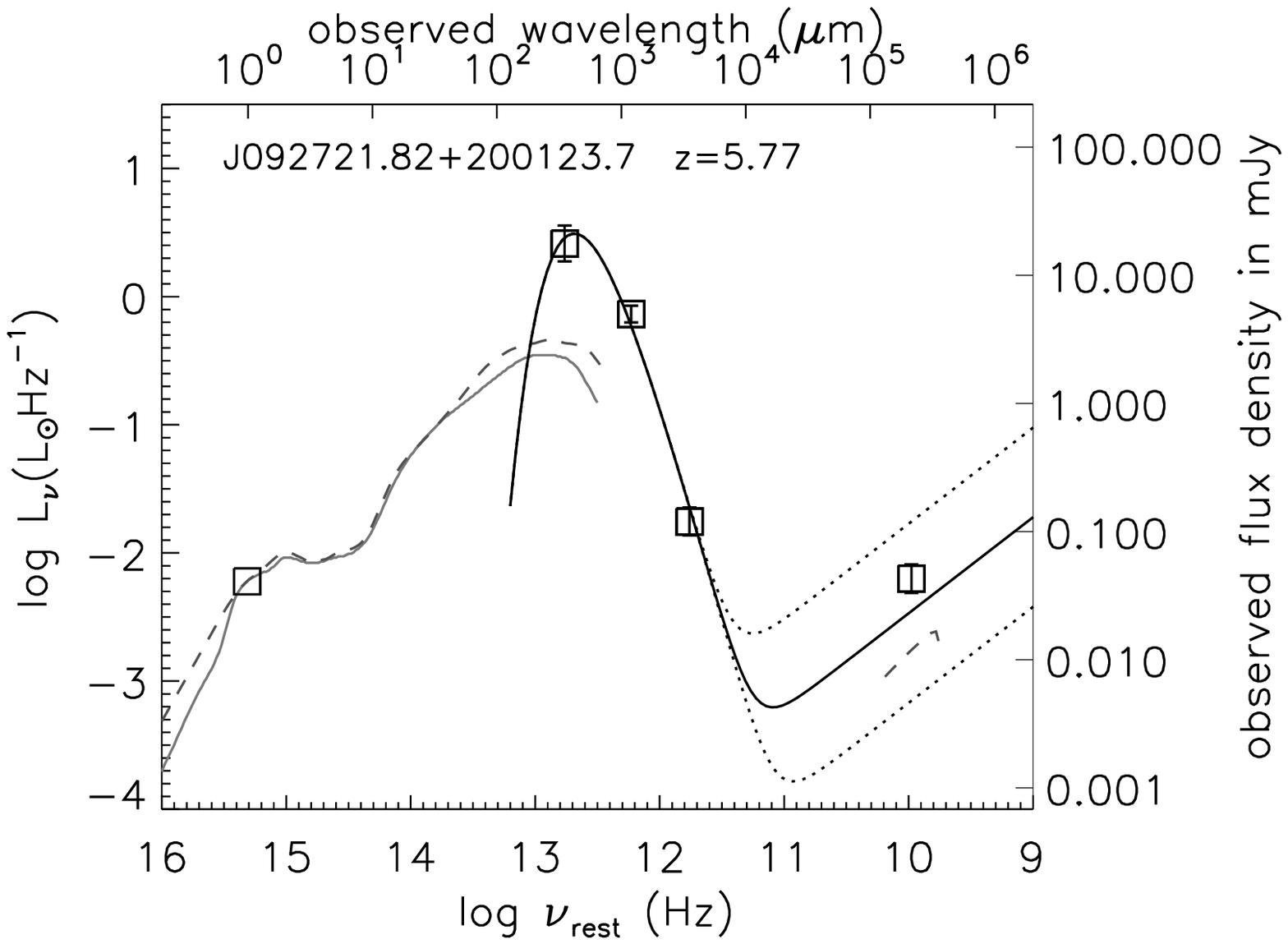}\\
\caption{Individual UV to radio SEDs of the 250 GHz detected
z$\sim$6 quasars. Squares show the observational data at 1450$\rm \AA$, 
3.6 $\rm \mu$m, 4.5 $\rm \mu$m, 5.6 $\rm \mu$m, 8.0 $\rm \mu$m, 
350 $\rm \mu$m, 450 $\rm \mu$m, 850 $\rm \mu$m, 1.2 mm (250 GHz), 
and 20 cm (1.4 GHz), and arrows denote 3$\rm \sigma$
upper limits. The 3 mm continuum data for the sources J0927+2001 
and J2054-0005 from the PdBI CO observations (Carilli et al. 2007; Wang et al. 2008b, in prep.) are also included.
Local quasar templates are plotted as in Figure-2
and normalized to rest frame 1450$\rm \AA$. The thick solid
line is a warm dust model (see Section 3.2) normalized to the
submm data and extended to the radio band
with the typical radio-FIR correlation of star forming galaxies, i.e.
q=2.34 (Yun et al. 2001), and the dotted lines factors of 5 
excesses above and below the typical q value.}
\end{figure}
\begin{figure}
\figurenum{3}
\plottwo{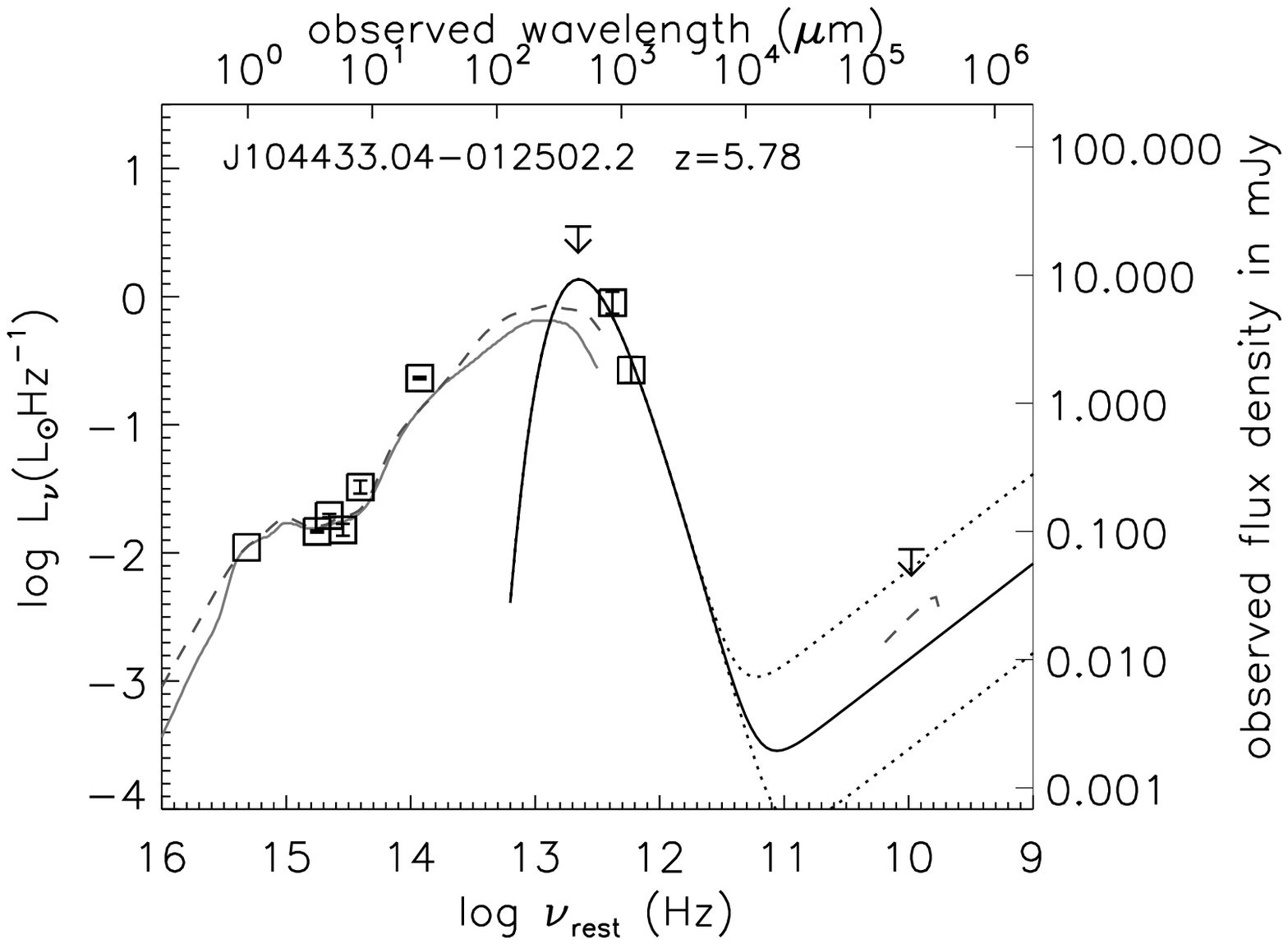}{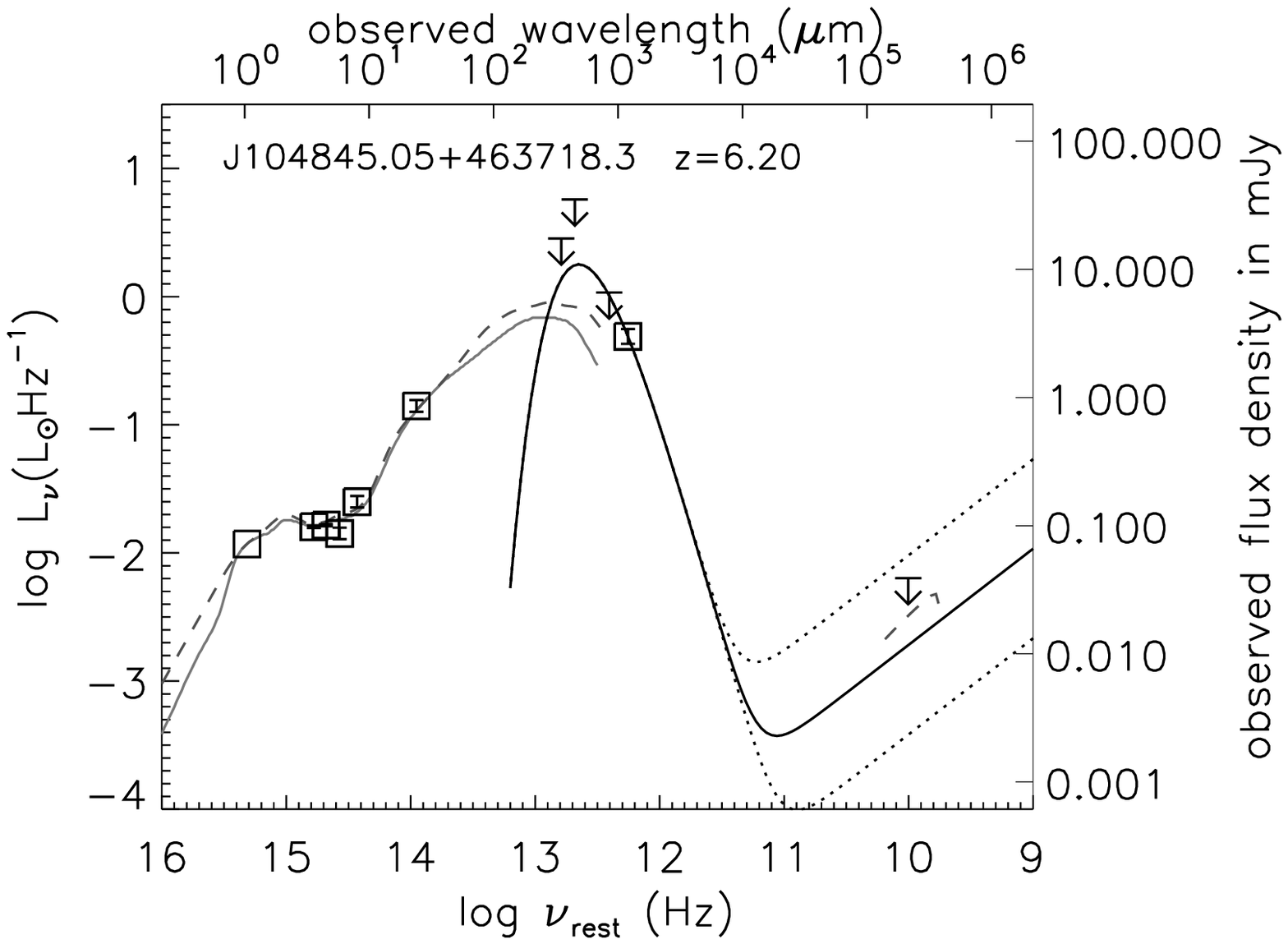}\\
\plottwo{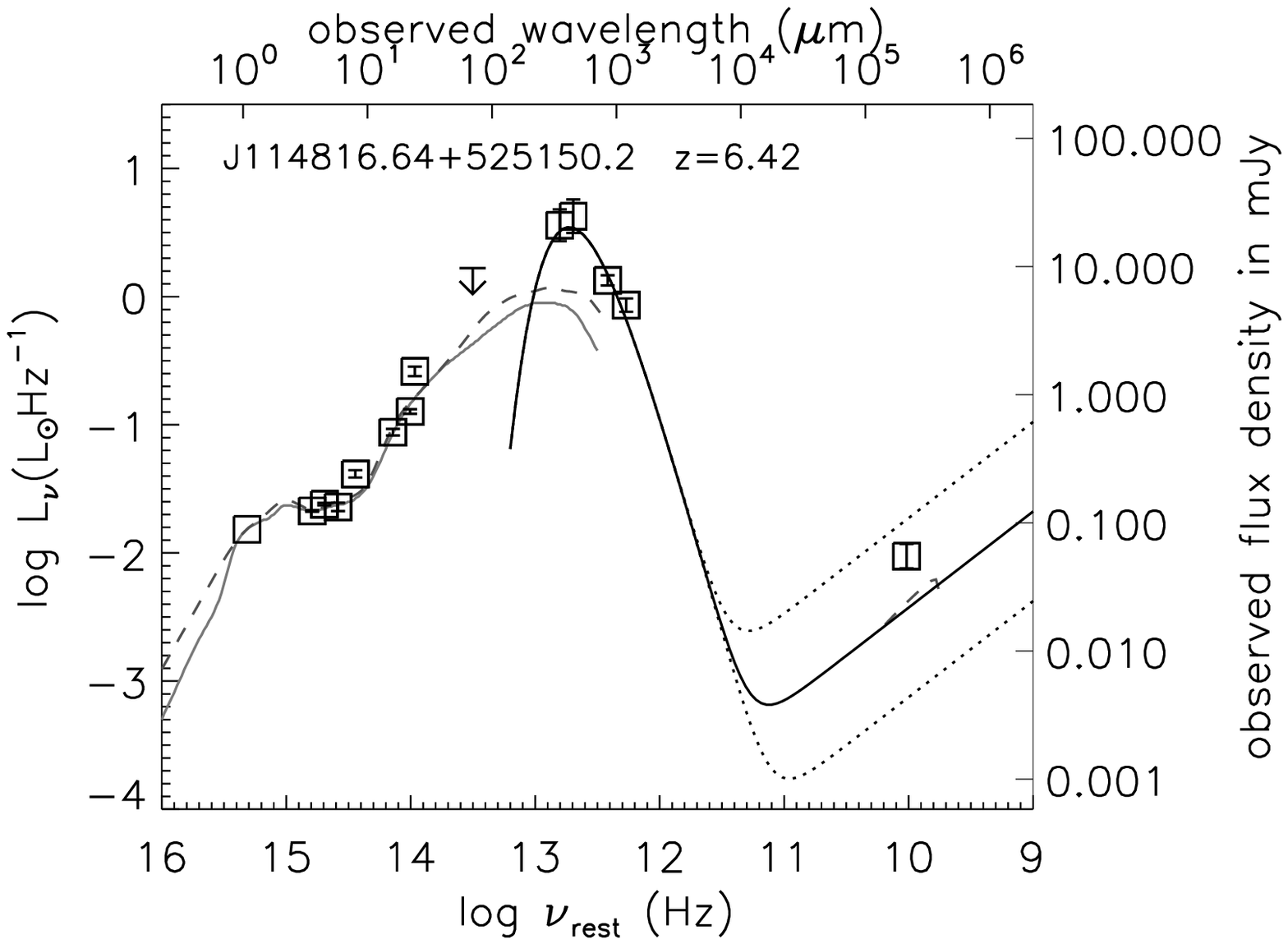}{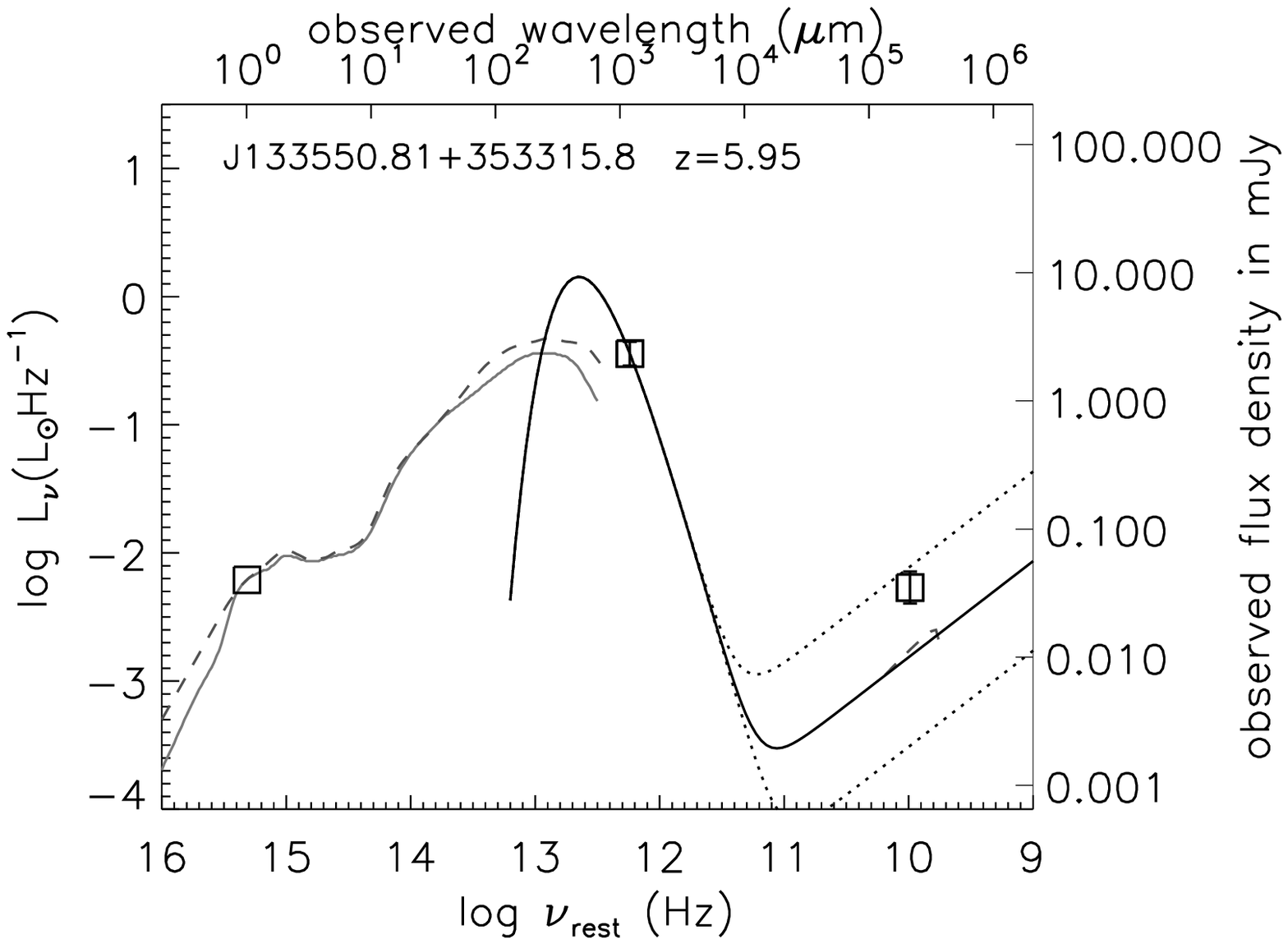}\\
\plottwo{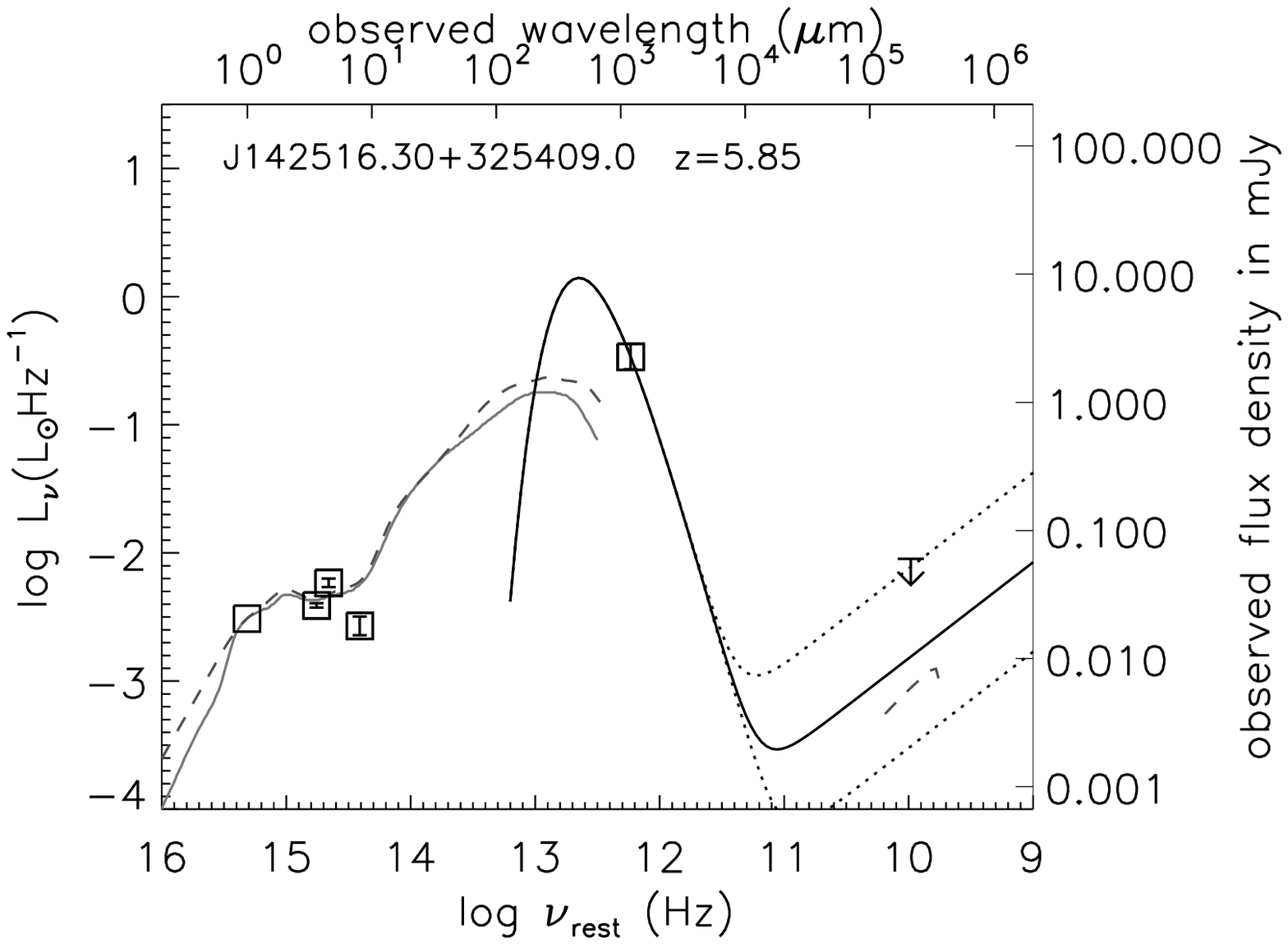}{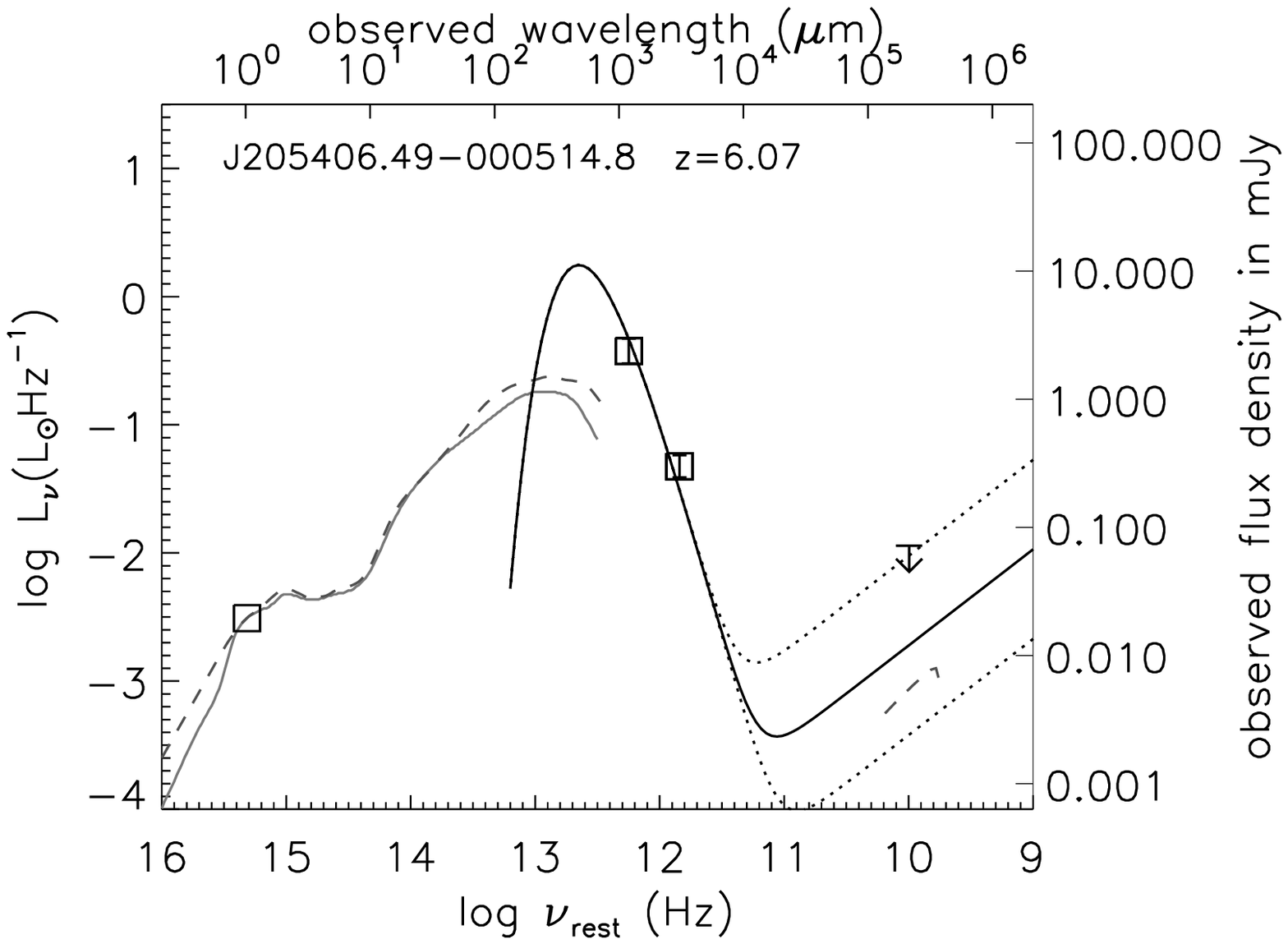}\\
\caption{Continued}
\end{figure}
\begin{figure}
\plotone{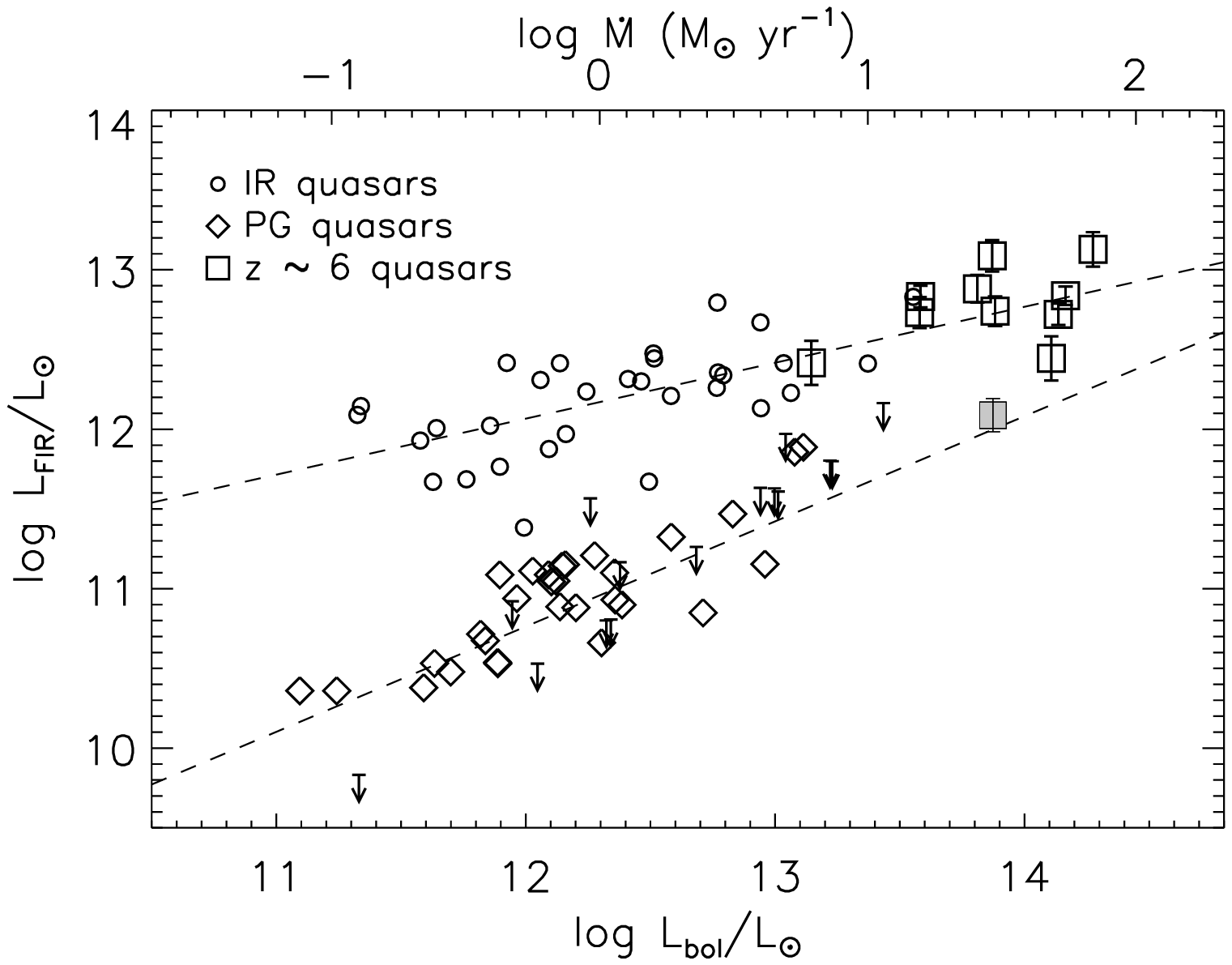}\\
\caption{The $L_{FIR}$--$L_{bol}$ correlation. The 250 GHz
detected z$\sim$6 quasars are plotted as black open squares with error
bars denoting 1$\sigma$ rms. The average value of the 250 GHz 
non-detections is plotted as a gray square. 
The local IR and PG quasars from Hao
et al. (2005) are plotted as circles and diamonds with
arrows denoting upper limits in $\rm L_{FIR}$. The dashed lines
represent the linear regression results for the two local quasar
samples. We convert $\rm L_{bol}$ to the mass accretion 
rate ($\rm \dot M$) at the top abscissa.}
\end{figure}

\begin{figure}
\plotone{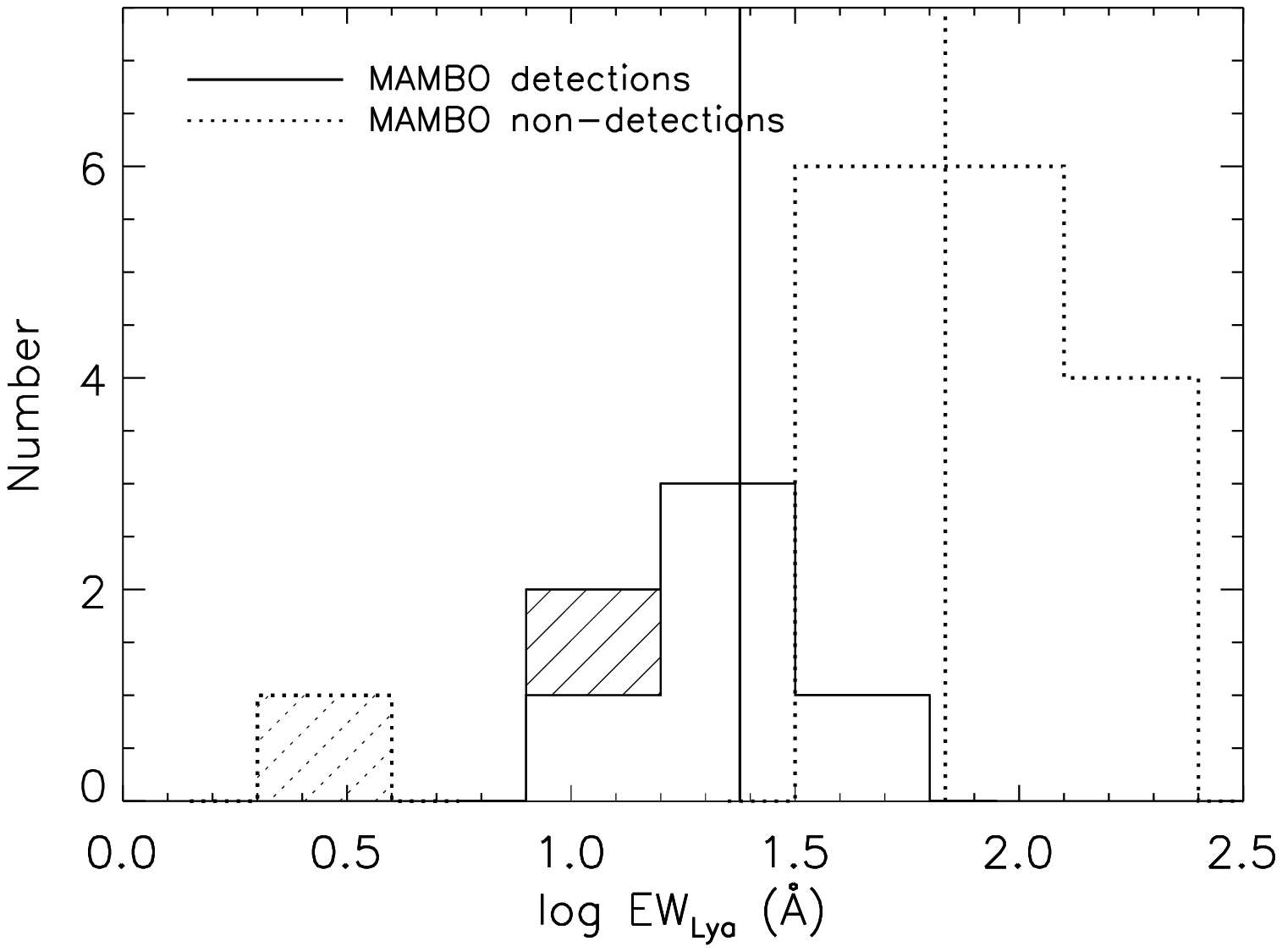}\\
\caption{The distribution of rest-frame equivalent widths 
of $\rm Ly\alpha$ (EW$\rm _{ly\alpha}$) for MAMBO detected (solid line)
and non-detected (dotted line) quasars at z$\sim$6. The shadowed 
area denotes the two sources that only have upper limits 
for the $\rm Ly\alpha$ emission. The solid and dotted lines 
represent the median values for MAMBO detections 
and non-detections, respectively.}
\end{figure}

\begin{figure}
\plotone{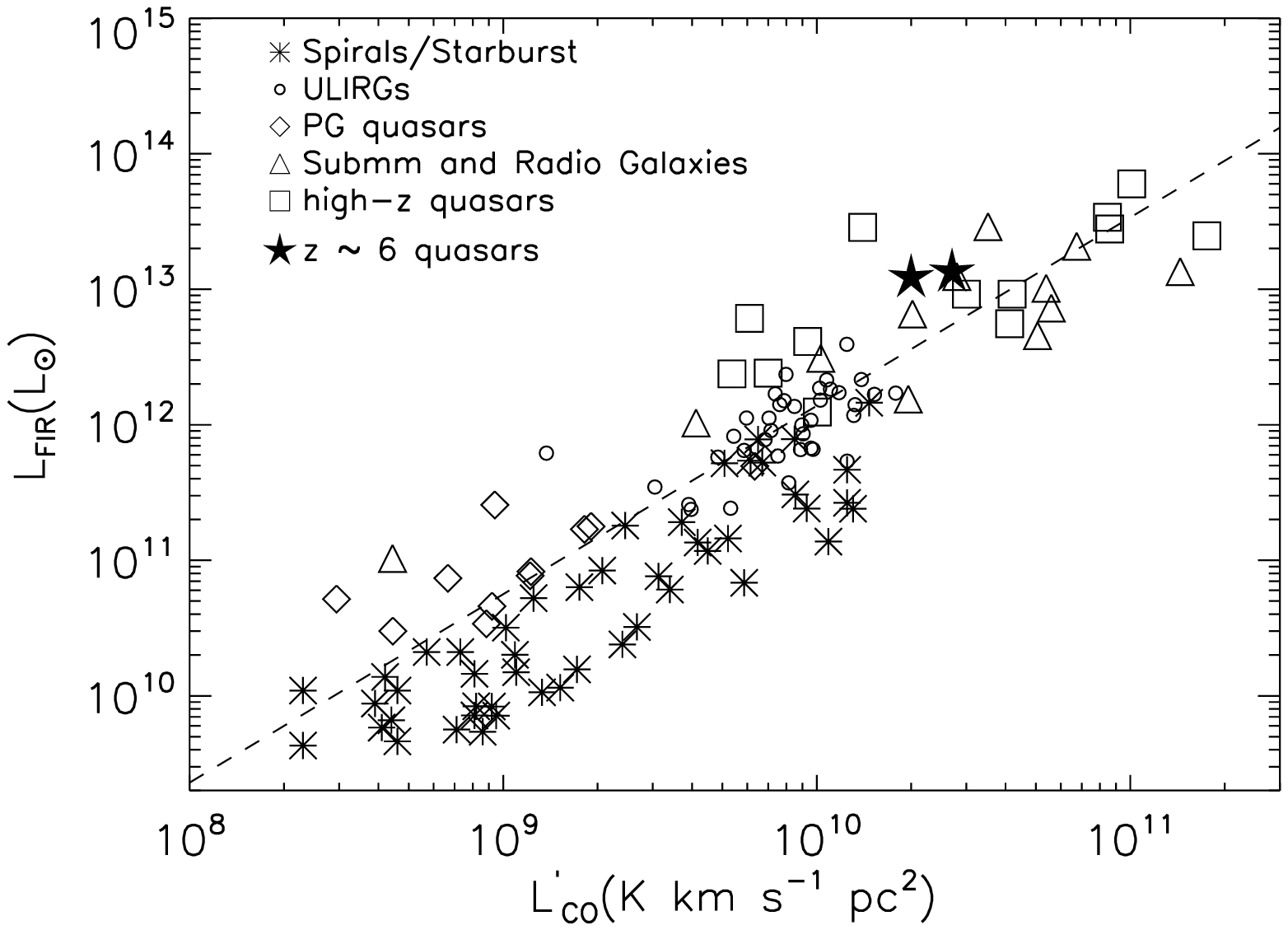}\\
\caption{$\rm L_{FIR}$ vs. CO luminosity ($\rm L'_{CO}$) for 
galaxies and quasars at various redshifts taken from Riechers et al. (2006).
The dashed line represents the relationship 
$\rm L_{FIR}\propto {L'_{CO}}^{1.39}$, derived
from low-z ULIRGs, starbursts, high-z submm and radio galaxies,
and low-z and high-z CO detected quasars in Riechers et al. (2006).
The stars represent the two CO detected quasars in the z$\sim$6 sample.}
\end{figure}

\end{document}